\documentclass[usenatbib]{mn2e}

\usepackage{url, amssymb, textcomp, verbatim, graphicx, times,float}
\usepackage[usenames, dvipsnames]{color}
\usepackage{lineno}
\newcommand{\adeg}[1]{{#1}$^{\circ}$}
\newcommand{\amin}[1]{{#1}$^\prime$}
\newcommand{\asec}[1]{{#1}$^{\prime\prime}$}
\newcommand{\thour}[1]{{#1}$^{\mathrm{h}}$}
\newcommand{\tmin}[1]{{#1}$^{\mathrm{m}}$}
\newcommand{\tsec}[1]{{#1}$^{\mathrm{s}}$}

\newcommand{\mjybeam}[1]{{#1}\,mJy\,beam$^{-1}$}
\newcommand{\mujybeam}[1]{{#1}\,$\mu$Jy\,beam$^{-1}$}
\newcommand{\mujy}[1]{{#1}\,$\mu$Jy}
\newcommand{\hms}[3]{\thour{#1}\tmin{#2}\tsec{#3}}
\newcommand{\dms}[3]{\adeg{#1}\amin{#2}\asec{#3}}
\newcommand{\sbeam}[2]{\asec{#1}$\times$\asec{#2}}

\newcommand{\fermi}{\emph{Fermi} }

\newcommand{\wise}{\emph{WISE} }

\newcommand{\atca}{\emph{ATCA} }

\def\aj{AJ}% % Astronomical Journal
\def\apj{ApJ}% % Astrophysical Journal
\def\apjl{ApJ}% % Astrophysical Journal, Letters
\def\apjs{ApJS}% % Astrophysical Journal, Supplement
\def\aap{A\&A}% % Astronomy and Astrophysics
\def\aaps{A\&AS}% % A&A Supplements
\def\aapr{A\&A~Rev.}% % Astronomy and Astrophysics Reviews
\def\baas{BAAS}% % Bulletin of the AAS
\def\mnras{MNRAS}% % Monthly Notices of the RAS
\title{An image-based search for pulsars among \textit{Fermi} unassociated LAT sources}

\author[Frail et al.]
{D. A. Frail$^1$, P. S. Ray$^{2}$, K.~P.~Mooley$^{3,10}$, P. Hancock$^{4,5}$, T. H. Burnett$^{6}$,
\newauthor{P. Jagannathan$^{1,7}$, E. C. Ferrara$^{8}$, H. T. Intema$^{9}$, F. de Gasperin$^{9}$,}
\newauthor{P. B. Demorest$^{1}$, K. Stovall$^{1}$ and M. M. McKinnon$^{1}$}
\\
  $^{1}$ National Radio Astronomy Observatory, 1003 Lopezville Road, Socorro, NM 87801, USA; Email: dfrail@nrao.edu\\
  $^{2}$ Space Science Division, Naval Research Laboratory, Washington, DC 20375-5352, USA \\
  $^{3}$ Astrophysics, Department of Physics, University of Oxford, Keble Road, Oxford OX1 3RH, UK \\
  $^{4}$ International Centre for Radio Astronomy Research (ICRAR), Curtin University, Bentley, WA 6102, Australia\\
  $^{5}$ ARC Centre of Excellence for All-sky Astrophysics (CAASTRO), Sydney, Australia\\
  $^{6}$ Department of Physics, University of Washington, Seattle, WA 98195-1560, USA\\
  $^{7}$ Department of Astronomy, University of Cape Town, Private Bag X3, Rondebosch 7701, Republic of South Africa \\
  $^{8}$ Center for Research and Exploration in Space Science and Technology and X-Ray Astrophysics Laboratory, \\
NASA Goddard Space Flight Center, Code 662, Greenbelt, MD 20771, USA\\
  $^{9}$Leiden Observatory, Leiden University, Niels Bohrweg 2, NL-2333CA, Leiden, The Netherlands\\
  $^{10}$ Hintze Research Fellow
}
\begin{document}
\maketitle

\begin{abstract}
We describe an image-based method that uses two radio criteria, compactness and spectral index, to identify promising pulsar candidates among \fermi Large Area Telescope (LAT) unassociated sources. These criteria are applied to those radio sources from the Giant Metrewave Radio Telescope all-sky survey at 150 MHz (TGSS ADR1) found within the error ellipses of unassociated sources from the 3FGL catalog and a preliminary source list based on 7 years of LAT data. After follow-up interferometric observations to identify extended or variable sources, a list of 16 compact, steep-spectrum candidates is generated. An ongoing search for pulsations in these candidates, in gamma rays and radio, has found six millisecond pulsars and one normal pulsar. A comparison of this method with existing selection criteria based on gamma-ray spectral and variability properties suggests that the pulsar discovery space using \fermi may be larger than previously thought. Radio imaging is a hitherto underutilized source selection method that can be used, as with other multi-wavelength techniques, in the search for \fermi pulsars.

\end{abstract}

 %  one of which had been discovered independently earlier


\section{Introduction}

One of the more significant contributions from the \textit{Fermi Gamma-ray Space Telescope} has been the discovery of large numbers of rotation-powered pulsars \citep{aaa+13}. Prior to launch, gamma-ray emission was expected to be detected among the normal, non-recycled radio pulsars but equal numbers of radio-quiet gamma-ray pulsars (``Geminga-like'') have been discovered \citep{car14}. More surprisingly, nearly half of the current sample of $\sim$200 gamma-ray pulsars are radio-loud millisecond pulsars (MSPs). Comparative studies of these populations have led to new understanding on the geometry and emission processes in neutron star magnetospheres \citep{har13}. \fermi MSPs are being used in the quest to detect the stochastic gravitational wave background \citep{dfg+13,brd15}. Among the MSP sample are significant numbers of exotic binaries known as ``redbacks'' and ``black widows'', representing an evolutionary phase when the system emerges from an earlier accretion-powered low-mass X-ray binary (LMXB) state to a rotation-powered radio MSP that is ablating its companion star. Rarer still are those ``transition'' systems that are seen to flip between the LMXB and MSP states \citep{sah+14,jrr+15,bh15}. Young, energetic pulsars have also been seen to undergo a type of mode-changing with accompanying changes in the pulse profile, the high-energy flux, and spin-down power \citep{abb+13}.

Since the release of the 4-year LAT catalog \citep[hereafter 3FGL;][]{aaa+15a}, the Large Area Telescope (LAT) continues to survey the gamma-ray sky finding an increasing number of sources \citep{mtf15}. By searching among the \fermi LAT unassociated sources, the population of pulsars has continued to grow. Most gamma-ray pulsars discovered to date have been found through direct pulsation searches at radio or gamma-ray wavelengths. Radio searches have been particularly effective; target selection based on gamma-ray properties is more efficient than blind sky surveys \citep[e.g.][]{ckr+15,cck+16}. The ephemerides obtained by timing these radio pulsars have in turn enabled additional detections of gamma-ray pulsations.

Increasingly, multi-wavelength imaging approaches are being used to complement these traditional pulsation search methods. X-ray counterparts are generally expected for both young, rotation-powered and millisecond pulsars \citep{ado+13,mmd+15,pb15}. Follow-up optical spectroscopy and photometry has revealed pulsars (or pulsar candidates) in binaries whose radio emission is suppressed in whole or in part by eclipses \citep{rfs+12,brr+13,kjy+14,scc+15}. 

The use of radio imaging to identify pulsar candidates has not been as widespread in the \fermi era \citep[see][]{rrb+15}. Historically this method was used to identify the first isolated millisecond pulsar and the first globular cluster pulsar  \citep{eric80,hhb85}. In two recent papers we have applied this technique to search for pulsars toward the \fermi 3FGL unassociated sources and toward the \fermi GeV excess \citep{fmji16,bdf+17}. We continue this work here using a larger sample of LAT sources from an all-sky analysis of 7 years of \fermi survey data, to identify further pulsar candidates. We have also undertaken higher angular resolution interferometry of the new candidates and the earlier 3FGL sample in an attempt to further strengthen our pulsar classifications by eliminating false positives within the sample, and we have carried out some preliminary searches for pulsations. In \S\ref{method} we describe our search method and apply it to the new source list while in \S\ref{obs} we describe the interferometric observations. The results of the search and the interferometric follow-up are given in \S\ref{sec:3fgl-results} and \S\ref{sec:7yr-results}, while in \S\ref{pulse} we summarize the search for pulsations. In \S\ref{discuss} we discuss the advantages and limitations of imaging approaches.

\section{Method}\label{method}

Our method for identifying promising candidates within the error ellipses of \fermi LAT unassociated sources relies on the fact that pulsars are compact, steep-spectrum radio sources. Pulsars stand out at low radio frequencies since they have steep power-law spectra typically from 100 MHz to several GHz, with negative slopes $\alpha=-1.8\pm0.2$ \citep[where S$_\nu\propto\nu^\alpha$;][]{mkk00}. After correcting for observational biases the {\it true} distribution of spectral indices has a mean $\alpha=-1.4\pm{1}$ \citep{blv13}.

Initial radio sources come from the GMRT 150 MHz All-Sky Radio Survey \citep[TGSS ADR1;][]{ijmf16}. This survey imaged the whole sky north of $\delta=-53^\circ$, covering 36,930 square degrees. The angular resolution is \sbeam{25}{25} for $\delta>$\adeg{19} and \sbeam{25}{25}$/\cos{(\delta-19^{\circ})}$ for more southern declinations. The median rms noise of the TGSS ADR1 images is \mjybeam{3.5} and the final 7$\sigma$ catalog contains 0.62 million radio sources. The source catalog and the processed images are publicly available on-line\footnote{\url{http://tgssadr.strw.leidenuniv.nl/doku.php}}.

The TGSS ADR1 is notable for its low noise and high angular resolution relative to other low frequency surveys \citep{heald2015,hch+17}. An all-sky centimetre survey such as the 1.4 GHz NRAO VLA Sky Survey \citep[NVSS;][]{ccg+98} has lower rms noise (\mjybeam{0.5}), but TGSS ADR1 sensitivity is superior for steep-spectrum sources, i.e. $\alpha\leq{-0.87}$. These properties are what make the TGSS ADR1 particularly well-suited for finding pulsars, both known and unknown. To illustrate this capability we note that phase-averaged emission was detected toward nearly 300 {\it known} pulsars in the TGSS ADR1 \citep{fjmi16}, more than all previous imaging surveys combined.

\begin{figure}
\includegraphics[width=\columnwidth]{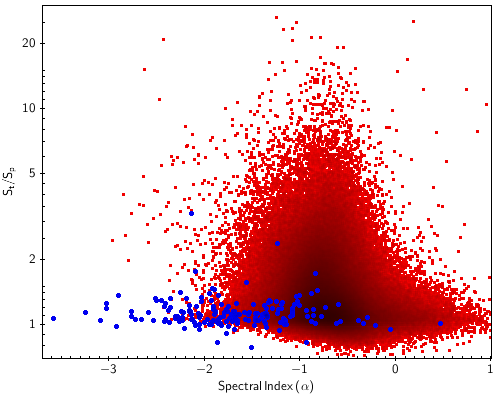}
\caption{The distribution of spectral indices versus compactness for background radio sources (red squares) and known pulsars (blue circles). Spectral indices are two-point values computed from the TGSS ADR1 and NVSS catalogs at 150 MHz and 1.4 GHz, respectively \citep{ijmf16}.  Compactness is defined as the ratio of the flux density (S$_t$) versus the peak brightness (S$_p$) at 150 MHz.}
\label{fig:criteria}
\end{figure}

To identify pulsar candidates it is necessary to define suitable spectral and angular criteria to distinguish compact, steep-spectrum sources from the large sample of radio sources. We begin by deriving a two-point spectral index (or limit) by comparing the TGSS flux density with the NVSS above $\delta\geq-40^\circ$ and the 843 MHz Sydney University Molonglo Sky Survey \citep[SUMSS;][]{bls+99} for $\delta<-40^\circ$. In Figure \ref{fig:criteria} we have plotted the (two-point) spectral index of a global source match of the TGSS and NVSS catalogs, versus a measure of source compactness, defined here as the ratio of the flux density (S$_t$) versus the peak intensity (S$_p$) at 150 MHz. The peak intensity is read directly from the images, while the flux density is derived from a Gaussian fit of the source. On the same plot we show all of the {\it known} pulsars for which we detected phase-averaged emission from the TGSS ADR1 at 150 MHz \citep{fjmi16}. The distribution of spectral indices for background radio sources is markedly different than that for pulsars. A typical radio source has a mean $\bar{\alpha}=-0.73$ \citep{ijmf16}, while the observed median distribution of pulsar spectral indices is closer to $-1.8$. Only ~0.3\% of all radio sources have $\alpha<-1.5$ \citep{ijmf16}, while 2/3 of all known pulsars have such steep spectral indices.

As noted above,  we use the ratio of the flux density of a source versus its peak intensity as our compactness criterion. \citet{fjmi16} argued that S$_t$/S$_p\lesssim{1.5}$ was sufficient to capture most viable pulsar candidates. Note that a more sophisticated measure of radio source compactness is defined in \citet{ijmf16} takes into account the increasing spread of this ratio with declining signal-to-noise. Such a measure is not particularly well suited for pulsars since their intensity can vary (due to scintillation) during an observation, violating the stationarity assumption that underpins radio interferometric imaging. Outliers with large S$_t$/S$_p$ ratios in Figure \ref{fig:criteria} include the bright and strongly scintillating PSR\,B1937+21 and pulsars whose flux densities are influenced by an underlying pulsar wind nebula.

An initial list of pulsar candidates selected on the basis of the two criteria of compactness (S$_t$/S$_p\lesssim{1.5}$) and spectral index ($\alpha\lesssim{-1.5}$) will still have too many false positives to be useful for conducting an efficient follow-up pulsation search. A list of possible false positives includes high redshift radio galaxies, variable or transient radio sources, cross survey calibration errors, and image or catalog artifacts. We discuss each of these in turn below. 

%Known pulsars are the easiest false positives to identify since the majority can be found in the latest version of PSRCAT \citep{mhth05}.  However, 
Not all steep-spectrum sources are pulsars. There is a population of steep-spectrum, luminous high redshift galaxies \citep[HzRGs;][]{md08} that can be mistaken for pulsars. HzRGs are interesting in their own right. They are useful signposts for identifying proto-clusters, and a suitably bright, highly redshifted 150 MHz radio source could be used to measure neutral hydrogen absorption of the intergalactic medium prior to the epoch of reionization \citep{cgo02}. The areal density of HzRGs is not well known but they can be separated from pulsars since the former have kpc-size extended structure while the later are point-like. The angular resolution of the TGSS ADR1 ($\sim$\asec{25}) is not sufficient to identify most HzRGs so follow-up observations are required at arcsecond resolution (see \S\ref{obs}). As an added benefit the arcsecond localization helps with identifying any optical and X-ray counterparts and reduces the pulsation search space. A detailed search of counterparts is beyond the scope of this paper.

Another class of false positives can be created as a result of measuring an initial spectral index from two all-sky catalogs taken at very different epochs. The NVSS was observed in circa 1995 while GMRT 150 MHz survey was carried out 16 years later (mean epoch=18 January 2011). In such a case short-lived transients or extreme variables could result in anomalously steep-spectrum sources. We suspect that transients are unlikely to be a serious contaminant as the radio sky is remarkably quiet at metre and centimetre wavelengths. The transient surface density at 150 MHz at the flux density threshold of the TGSS ADR1 is estimated to be 6.2$\times 10^{-5}$ deg$^{-2}$  by \citet{mkc+17}. Transient rates at 1.4 GHz at the sensitivity level of the NVSS are similarly low \citep[see][and references therein]{mhb+16}.
 
Variable radio sources can be a more significant source of false positives. Pulsars can be variable sources but in imaging mode interferometric surveys tend to average out variability timescales of minutes to hours \citep{fjmi16}. Although the study of low-frequency variability at the sub-Jansky level has just begun, initial results suggest that approximately 1.5\% of radio sources  at 150 MHz are ``strong variables'', defined roughly as sources whose max-to-min flux density difference varies by 1.4 times the average flux density on timescales of a year or more \citep{mool18}. Similar numbers are found for variability at 1.4 GHz \citep{hdb+16}. Variability of this magnitude can produce changes in the mean spectral index of order $\pm$0.15. This amounts to a simple error term when studying the bulk of the radio source population, but the tails of the spectral index distribution are more susceptible. Even though only a fraction of radio sources show this level of variability, the smaller number of sources with very negative ($\alpha<-1.5$) or positive power-law spectral slopes will be contaminated by the more numerous flatter spectral index sources scattering into the tail. We estimated the magnitude of the effect by taking a pessimistic case where {\it all} the variables lie in a narrow range of $-1.50<\alpha<-1.25$ and 1.5\% of these sources vary by enough to scatter into  $\alpha\leq{-1.50}$. In this case as much as 5\% of the steep-spectrum candidates could be due to variables. Variability does not have to be intrinsic to the source either. Direction-dependent flux density calibration errors between surveys mimic the effects of variability, again having a bigger effect at the tails of the spectral index distribution.

There are also false positives that originate as image or catalog artifacts. These are the result of using sky surveys taken by different telescopes with different rms noise sensitivity, varying sensitivity to spatial structure, and different angular resolution. Additionally, source-finding algorithms used to generate the catalogs have very different levels of reliability and completeness. There has also been a gradual improvement in these algorithms with time \citep{hmg+12,mfo+13}. For example, we have found that the SUMSS catalog is missing substantial numbers of radio sources that are clearly visible in the SUMSS images. Visual inspection of the NVSS images can also reveal weak radio sources not in the original catalog. Image artifacts also appear in all three of the surveys that we used. While the TGSS ADR1 catalog has remarkably good source reliability \citep[see Fig. 18 of][]{ijmf16}, there are still false source identifications of sidelobe structures, caused by residual calibration errors near bright radio sources. The short snapshots and compact configuration used to carry out the NVSS make it less reliable for point-source identification in the Galactic plane. All of these false positives can be easily picked out by visually inspecting the candidates in the original images.

Summarizing the above discussion, we have shown that a promising list of pulsar candidates can be generated by identifying compact (S$_t$/S$_p\lesssim{1.5}$) and steep-spectrum ($\alpha\lesssim{-1.5}$) radio sources from a low frequency catalog (Fig. \ref{fig:criteria}). However, prior to undertaking a resource-intensive pulsation search, the reliability of each pulsar candidate must be verified by identifying and removing false positives from the sample. In some cases (e.g. missing sources, artifacts) all that is required is visual inspection of the original images. When possible we use other existing survey data as a check on variability and to improve on the two-point spectral index values. This includes the Faint Images of the Radio Sky at Twenty centimeters survey \citep[FIRST;][]{bwh95} at 1.4 GHz, the Westerbork Northern Sky Survey  \citep[WENSS;][]{rtb+97} at 325 MHz, the Very Large Array Low-frequency Sky Survey Redux \citep[VLSSr;][]{lcv+14} at 74 MHz, and the recently released GLEAM survey from 72 to 231 MHz \citep{hch+17}. Follow-up interferometric observations remain an important final step to identify extended sources such as HzRGs, to provide further constraints on variability, and to improve on the accuracy of the spectral index and sky position.

%makes it difficult to automate of provide reliable ststistics 
% polarization. Can help if you have it. 

%Relative to the extragalactic sources that dominate the source counts of the radio sky, steep spectrum sources are rare \citep{ki08}, and yet 

\subsection{The 3FGL Sources}\label{meth:3fgl}

In a recent paper, we used this radio image-based method to identify promising pulsar candidates among the unassociated sources in the 4-year 3FGL catalog \citep[hereafter Paper I;][]{fmji16}. We searched the error ellipses of the 839 \fermi sources 
above the southern declination limit of the TGSS ADR1 of $\delta=-53^\circ$ finding 1485 radio sources at 150 MHz. From an initial candidate list of 25 steep-spectrum sources we used existing radio images and catalogs to generate 11 point-like, steep-spectrum pulsar candidates. This work was carried out before \citet{fjmi16}, and so a less stringent compactness criterion was used than that adopted here (i.e. S$_t$/S$_p\lesssim{1.5}$). This approach left open the possibility that our candidate list from Paper I had significant contamination from HzRG. Thus we have undertaken interferometric observations of most of the 3FGL candidates (see \S\ref{obs}).

\subsection{The 7-year \emph{Fermi} LAT Sources}\label{meth:7year}

For the current work we used an all-sky source list based on 7 years of LAT survey data, using the updated Pass 8 data set and the same likelihood-based procedure to detect point sources used for the 3FGL catalog \citep{aaa+15a}, but going down to lower significance than the regular LAT catalogs. This analysis was a preliminary version of the procedure that will be used to produce the next public release LAT source catalog. It is expected that the next public release of the LAT source catalog will use an eight year data set. Because our source list does not correspond to a particular \fermi LAT catalog, we give them generic source names like \textsc{Fermi} JHHMM.m+DDMM, in accordance with IAU naming conventions. Our source list
%\footnote{P302\_7years\_uw982\_assoc\_v7r1.fits}
contains 7091 discrete gamma-ray sources, of which 2652 are unassociated sources. The term ``unassociated'' in the \fermi context means a gamma-ray source that is lacking a reliable association with sources detected at other wavelength bands. While the number of unassociated sources is more than twice that in the 3FGL catalog, the fraction relative to the total number of discrete sources is still about one third. Above the de`<clination cutoff of our radio source search ($\delta\geq -53^\circ$) there are 2313 unassociated sources. The mean semi-major and semi-minor axes (95\% confidence) for these unassociated sources is \adeg{0.16} and \adeg{0.12}, respectively, and the total search area is 182.2 square degrees.

Our initial radio source matching with \fermi unassociated error ellipses was done with the GMRT 150 MHz All-Sky Radio Survey \citep{ijmf16}. We used the latest version of the TGSS $5\sigma$ catalog which contains 707,255 sources. This catalog is expected to be made public along with a new data release (TGSS ADR2) in 2018. Since the total TGSS area is about 36,930 square degrees (11.25 sr) the areal density of radio sources on the sky is 19.2 radio sources per square degree.

We used TOPCAT \citep{tay05} to search the $5\sigma$ catalog for all 150 MHz radio sources within the 95\% confidence error ellipses of all the unassociated LAT sources in our new source list. A total of 3866 radio sources were found toward the 2313 unassociated sources. Of these 2313 unassociated gamma-ray sources 1367 have one or more radio sources within their 95\% error ellipses, while another 946 have no radio sources to the limits of the TGSS ADR1 catalog. The radio source counts toward \fermi unassociated error ellipses show an excess of approximately 10\% over the background. We estimate the background radio source counts in two ways. The simplest method is to multiply the total area searched (182.2 square degrees) times the areal density of the TGSS 5$\sigma$ catalog (19.2 per square degrees) giving 3498 sources, for an excess of 368 radio sources. For more accurate values we must take into account the decrease of radio source counts in the Galactic plane.  For this we simulated populations of gamma-ray sources whose number and total area were the same as the 2313 unassociated sources but with random positions according to the prescription in \S{3.2} of \citet{aaa+11}. We generated 500 random gamma-ray source lists and carried out matched searches of radio sources for each. The average number of background radio sources is 3523.5 sources, distributed with a Gaussian of width $\pm$67.9. These two estimates agree within the errors. Thus the excess in TGSS ADR1 radio sources toward unassociated \fermi sources at 150 MHz is 342.5 sources, or 5$\sigma$.

%These two estimates differ slightly because many \fermi error ellipses lie in the Galactic plane where the increased brightness temperature, due to the diffuse synchrotron emission from our Galaxy, results in higher receiver noise temperature and hence a slight reduction in the radio source counts relative to the all-sky value.

A radio excess toward \fermi unassociated sources is unsurprising since the known population of \fermi sources are mostly active galactic nuclei and nearly all of these have a radio counterpart \citep{aaa+15b,mtf15}. Here we are concerned with identifying pulsar candidates and so we start by calculating two-point spectral indices for the 3866 TGSS sources at 150 MHz and using either NVSS at 1.4 GHz or SUMSS at 0.84 GHz. For the TGSS ADR1 sources with neither a NVSS nor a SUMSS counterpart, we calculated an upper limit on the spectral index based on the completeness limit of these surveys (NVSS=2.5 mJy, SUMSS=10 mJy). We next apply the compactness and spectral criteria. As noted above we used $\alpha\leq-1.5$ as our definition of a steep-spectrum radio source and we used S$_t$/S$_p\lesssim{1.5}$ to define compactness.  For the TGSS-NVSS comparison we found 6 compact, steep-spectrum matches and another 20 with only upper limits on $\alpha$ (i.e. TGSS detections without NVSS counterparts). For the SUMSS there were no steep-spectrum matches but there were 29 upper limits. From this initial candidate list of 55 steep-spectrum sources we used existing radio images and catalogs to identify false positives. Only two of 29 SUMSS candidates were genuine steep-spectrum candidates. The remaining sources appear to be imaging artifacts or are simply typical non-thermal radio sources that just failed to make it into the SUMSS catalog. There are similar issues with the NVSS and TGSS images but not of the same magnitude. After visual inspection of the TGSS-NVSS images, we found 10 candidates; 4 with spectral indices derived from both NVSS and TGSS and 6 upper limits with only TGSS catalog flux densities. Our final list prior to follow-up interferometry (\S\ref{obs}) consisted of 12 pulsar candidates.

\section{Interferometer Observations}\label{obs}

As noted in \S\ref{method}, interferometric follow-up observations have two main goals. The first is to identify false positives, i.e. high redshift galaxies that are resolved at arcsecond angular resolution. The second goal is to verify the flux density at centimeter wavelengths and, where possible, improve on the position to help counterpart and pulsation searches. Detailed results are discussed in \S\ref{sec:3fgl-results} and in \S\ref{sec:7yr-results}.

\subsection{VLA Observations}\label{obs:vla}

We obtained data with the Karl G. Jansky Very Large Array (VLA) as part of project 16A-460 using Director's Discretionary Time. Observations were carried out between May and July 2016 in CnB and B array configurations. Under the assumption that the spectra of the sources will be falling steeply with frequency, we chose S band (2--4 GHz) as a compromise between maintaining adequate continuum sensitivity for these steep-spectrum sources and requiring arcsecond angular resolution to resolve extended emission. On-source integration times were typically 20-30 minutes.  We used the standard Wideband Interferometric Digital Architecture (WIDAR) correlator setup for continuum observing with 16 spectral windows, 64 2-MHz wide channels each to get 2 GHz of total bandwidth centered on 3.0 GHz, and 5-sec integrations. RFI flagging and calibration of the data were done using the automated VLA CASA calibration pipeline\footnote{https://science.nrao.edu/facilities/vla/data-processing/pipeline} implemented for CASA 4.6. The target data were then split off and each target pointing was imaged up to the half-power point of the primary beam. Suitable pixel size was chosen in order to sample the synthesized beam with four pixels.

\subsection{ATCA Observations}\label{obs:atca}

We obtained data with the Australia Telescope Compact Array (ATCA) as part of project CX355 using Director's Discretionary Time. Observations were conducted on 28-29 April 2016 in the 6A (6 km) configuration and on 16-17 June 2016 in the EW352 (compact) configuration.  A full 2 GHz of bandwidth was correlated with the Compact Array Broadband Backend \citep[CABB;][]{wfa+11} centered on 2.1 GHz and using a channel width of 1 MHz. On-source integration times were typically 1 hour. PKS\,1934$-$638 was the primary flux density calibrator, while 0823$-$500 was used for delay and bandpass calibration. Phase calibrators were selected close to each target source. The visibility data were calibrated and imaged using the MIRIAD package following standard practices \citep{stw95}.

% make sure the AGN figure is put in the right place in the final submitted manusacript.
\begin{figure*}
\includegraphics[width=7.5in]{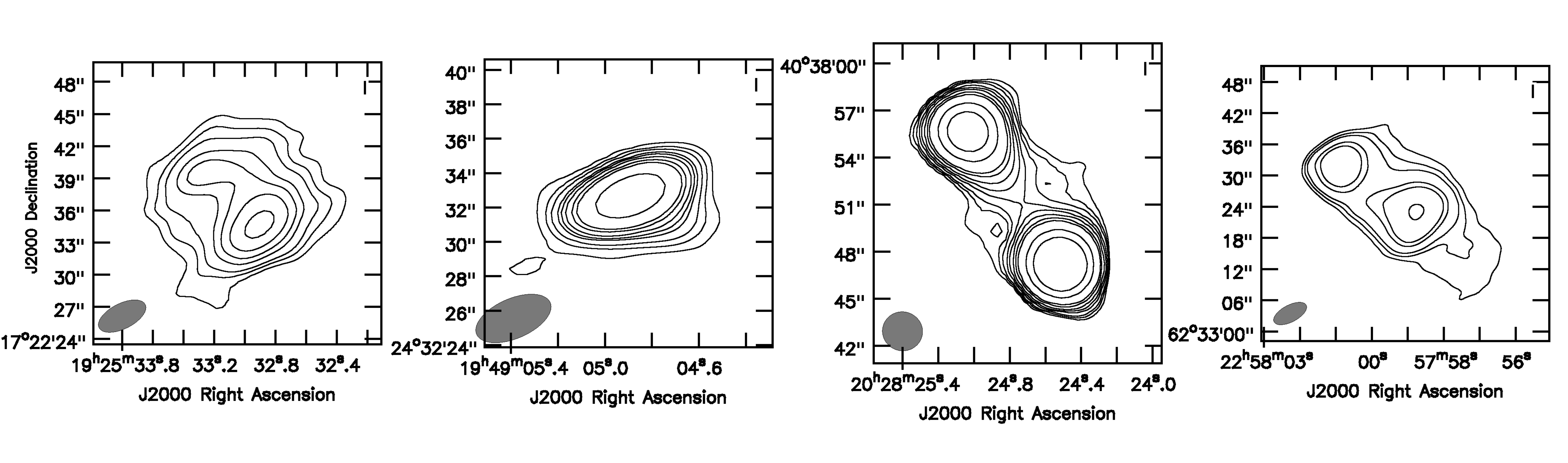}
\caption{Contour plots for steep-spectrum radio candidates observed by the VLA at 3.1 GHz (left to right: 3FGL J1925.4+1727, 3FGL J1949.3+2433, 3FGL J2028.5+4040c, \textsc{Fermi}\,J2259.1$+$6233) in Fermi error ellipses that are resolved by the VLA. Due to their extended nature, these are no longer considered pulsar candidates. The synthesized beam for each cutout is shown as a grey ellipse. The contour levels are at 3, 5, 7, 10, 12, 15, 20, 50 and 100 times the rms noise for the 3FGL sources, and 15, 20, 25, 40, 50, 100 times the rms noise for \textsc{Fermi}\,J2259.1$+$6233. The RMS noise in the respective images (left to right) is 48, 42, 44, and \mujybeam{23}.}
%The center frequencies for the images lie between 2.8 GHz and 3 GHz.} See \S\ref{obs:vla} and \S\ref{obs:atca} for details.}
\label{fig:agn}
\end{figure*}

\section{Results}\label{results}

\subsection{The 3FGL Candidate Sample}\label{sec:3fgl-results}

Interferometric observations were made with the VLA ({\S\ref{obs:vla}) and \atca (\S\ref{obs:atca}) for all but three of the original 11 steep-spectrum candidates from Paper I. Table \ref{tab_flux} lists all 11 3FGL sources from Paper I with their S-band flux densities (or upper limits) along with short notes. The three sources that were not observed have no S-band values in Table \ref{tab_flux}. Two of these three unobserved sources (3FGL J1827.6$-$0846 and J1901.5$-$0126) had been detected as gamma-ray MSPs in a blind search using the TGSS positions prior to the start of the VLA and \atca observing \citep[see \S\ref{pulse} and][]{c+18}, while the third source (3FGL J0258.9+0552) already had a good interferometric image from FIRST. For point sources or upper limits that remain viable pulsar candidates we list the interferometer that observed each field in column 3. If a source was subsequently determined to be an active galactic nucleus (AGN) or a known pulsar (PSR), we do not include the S-band flux density and we make a short-hand notation about the object to this effect in Column 3.

While the candidate pulsars toward 3FGL J1925.4+1727, J1949.3+2433 and J2028.5+4040c were point-like to the \asec{25} beam of the TGSS images, they appear fully resolved in the follow-up interferometric images with a resolution of ~\asec{4}. VLA images of each source are shown in Fig.\,\ref{fig:agn}. The first two sources show diffuse emission on a scale of \asec{6} to \asec{12}, while the third source is resolved into a double with a separation of \asec{10}. These three resolved sources are no longer considered compelling pulsar candidates. \citet{lsb+17} argue that PSR\,J1925+1720, a young, energetic pulsar that lies just outside of the nominal \fermi error ellipse is the high energy counterpart.

% I got the beam from the fits files that Paul sent us in dropbox. 13 by 4 arcsec with -11 deg CCW
The pulsar candidate within the 3FGL J1830.8$-$3136 is unresolved by the \atca at 2.1 GHz with a resolution of \asec{13}$\times$\asec{4}. The peak brightness is 1.86$\pm$\mjybeam{0.05} with an improved J2000 position of R.A.=\hms{18}{30}{38.73} and $\delta$=\dms{$-$31}{35}{7.1}, and an uncertainty of $\pm$\asec{0.2}. With this third detection we can refit the power-law slope and derive $\alpha=-1.46\pm{0.10}$. This spectral slope is flatter than its original value from Paper I. While this compact radio source within 3FGL J1830.8$-$3136 could still be a pulsar, it nominally lies outside of our cutoff criterion of $\alpha\leq{-1.5}$.

The VLA and \atca did not detect significant emission toward four candidates. The \atca 3$\sigma$ upper limits at 2.1 GHz for the TGSS radio sources toward 3FGL J1533.8$-$5231, J1639.4$-$5146 and J1747.0$-$3506 are \mujybeam{300}, \mujybeam{600}, and \mujybeam{240}, respectively. These upper limits are all consistent with an extrapolation of the original two-point spectral indices derived in Paper I and thus these remain pulsar candidates. The same cannot be said for the VLA non-detection of 3FGL J2210.1+5925. The 3$\sigma$ upper limit of \mujybeam{55} is well below the extrapolation of the original power-law spectrum that predicts \mujy{850}, based on four flux density measurements from 75 to 1400 MHz. The sharp drop appears to be real and may be a result of variability or a resolution effect. To explore the latter, we convolved the image with a \asec{25} beam to search for extended emission but none was found above the noise. While 3FGL\,J2210.1+5925 could still be a (variable) pulsar candidate,  a more conservative interpretation of the large S$_t$/S$_p$ ratio is that it is an extended radio source (see discussion in \S\ref{method}).

To summarize, of the original list of 11 pulsar candidates from Paper I and Table \ref{tab_flux}, six remain as promising candidates defined by our criteria of compactness and spectral slope. These sources are listed in Table \ref{tab_candi}. In column one we list the source name. Columns two, three, and four contain information about the pulsation searches in \S\ref{pulse}. They are discussed in more detail in that part of the paper but in order they are the type of search, a ``pulsar-ness'' ranking based on gamma-ray properties (see \S\ref{discuss}), and a summary of the properties of the new pulsars.

\begin{table}
\caption{Measured interferometer flux densities or peak brightness limits.\label{tab_flux}}
\begin{tabular}{lcr}
\hline
\emph{Fermi} Name & S-band & Notes \\
\hline
\multicolumn{3}{c}{3FGL Sources} \\
\hline
3FGL J0258.9$+$0552  & \omit & FIRST \\
3FGL J1533.8$-$5231  & $<0.1$ & ATCA \\
3FGL J1639.4$-$5146  & $<0.2$ & ATCA \\
3FGL J1747.0$-$3506  & $<0.08$ & ATCA \\
3FGL J1827.6$-$0846  & \omit & MSP \\
3FGL J1830.8$-$3136  & 1.86$\pm$0.05 & $\alpha>-1.5$ \\
3FGL J1901.5$-$0126  & \omit & MSP \\
3FGL J1925.4$+$1727  & \omit & AGN \\
3FGL J1949.3$+$2433  & \omit & AGN\\
3FGL J2028.5$+$4040c & \omit & AGN \\
3FGL J2210.1$+$5925 & $<0.018$ & Var.\\
\hline
\multicolumn{3}{c}{7-Year Sources} \\
\hline
\textsc{Fermi} J0628.6$+$0511 & 1.47$\pm$0.05 & VLA \\
\textsc{Fermi} J1236.5$+$1133 & 3.08$\pm$0.03 & VLA \\
\textsc{Fermi} J1555.6$-$2906 & $<0.017$      & VLA \\
\textsc{Fermi} J1646.5$-$4406 & 1.4$\pm$0.2   & ATCA \\
\textsc{Fermi} J1722.0$-$3204 & \omit    & PSR \\
\textsc{Fermi} J1728.1$-$1608 & $<0.012$ & VLA\\
\textsc{Fermi} J1739.3$-$2530 & $<0.023$ & VLA\\
\textsc{Fermi} J1803.1$+$1400 & $<0.024$ & VLA \\
\textsc{Fermi} J1833.0$-$3839 & $<0.3$   & ATCA\\
\textsc{Fermi} J1843.8$-$3834 & 4.14$\pm$0.13 & ATCA\\
\textsc{Fermi} J2000.8$-$0300 & 0.75$\pm$0.03 & VLA \\
\textsc{Fermi} J2259.1$+$6233 & \omit & AGN \\
\hline  
\multicolumn{3}{l}{Note: S-band detections in mJy, upper limits in mJy\,beam$^{-1}$.}
\end{tabular}
\end{table}

%are 3FGL sources J0258.9+0552, J1533.8$-$5231, J1639.4$-$5146 and J1747.0$-$3506, J1827.6$-$0846 and J1901.5$-$0126.

\subsection{The 7-Year \emph{Fermi} Candidate Sample}\label{sec:7yr-results}

In \S\ref{method} we described the method for identifying the original 12 pulsar candidates from the 7-year LAT source list (Table \ref{tab_flux}). Two of the 12 sources were eliminated from further consideration. Our steep-spectrum candidate toward \textsc{Fermi} J1722.0$-$3204 is a known pulsar PSR\,B1718$-$32 from PSRCAT \citep{mhth05}. This is an old, slow pulsar with a period of 477 ms with a spin-down energy of 2.3$\times 10^{32}$ erg s$^{-1}$. Given the low spin-down energy, this pulsar is likely not associated but rather is a line-of-sight coincidence with the gamma-ray source. A second steep-spectrum candidate toward \textsc{Fermi} J2259.1$+$6233 was resolved by the VLA and is therefore not a pulsar (see Fig.\ref{fig:agn}). We discuss the remaining 10 sources individually in the subsections below. For each of these candidates we searched for additional radio data to improve upon the two-point spectral indices, and we looked for multi-wavelength counterparts. Their properties are summarized in 
%Table \ref{tab_candi} and 
Table \ref{tab:steep}. Table \ref{tab:steep} lists the radio properties of these candidates taken directly from the TGSS ADR1 catalog, i.e. the J2000 right ascension (RA) and declination (DEC), Galactic longitude ($l$) and latitude ($b$), the flux density in mJy (S$_t$), and the peak brightness (S$_p$) in  mJy\,beam$^{-1}$. For the VLA and data from \S\ref{obs:vla} and \S\ref{obs:atca}, the peak intensity is read directly from the image, while the flux density is derived from a Gaussian fit of the source. The power-law spectral index for each source in Table \ref{tab:steep} is derived for each source in the sub-sections below. If more accurate positions were measured they are also given in the discussion subsection for individual sources below. 

In Table \ref{tab:gamma} we list the properties of the 7-year unassociated gamma-ray sources based on our analysis. We list the semi-major (Maj) and semi-minor (Min) axes of the 95\% confidence positional ellipses, the spectral type (Spec Type), the gamma-ray flux (F$_\gamma$), a $\gamma$-ray spectral index (Spec. Index), energy flux at 100 MeV (E$_{100}$), the test statistic (TS) of the source derived from the likelihood analysis, and  the angular offset of the centroid of the \fermi source and the radio source ($\Delta\theta$). The current list includes two sources with TS$<25$. Such sources are not released in standard catalogs since it is more likely that they are spurious.
%there is a significant probability that they are spurious.
 Table 1 of Paper I has a similar list of properties for the 3FGL sample.

\begin{table}
\caption{Compact, Steep-Spectrum Pulsar Candidates with Interferometric Imaging and Pulsation Searches. \label{tab_candi}}
\begin{tabular}{llll}
\hline			
Source Name & Pulse Search & $\gamma$ grade & PSR Notes \\
\hline
\multicolumn{4}{c}{3FGL Sources} \\		
\hline
3FGL\,J0258.9$+$0552  & $\gamma$, GBT & unlikely & \omit \\
3FGL\,J1533.8$-$5231  & \omit         & unlikely & \omit \\
3FGL\,J1639.4$-$5146  & $\gamma$      & psr-A     & \omit \\
3FGL\,J1747.0$-$3506  & $\gamma$, VLA & psr-D     & \omit \\
3FGL\,J1827.6$-$0846  & $\gamma$, GBT & unlikely & gm (2.2 ms)   \\
3FGL\,J1901.5$-$0126  & $\gamma$, GBT & psr-C     & grm (2.8 ms)   \\
\hline  
\multicolumn{4}{c}{7-Year Sources} \\		
\hline

J0628.6$+$0511 & Arecibo, VLA      & psr-D     & \omit \\
J1236.5$+$1133 & Arecibo  & unlikely & \omit \\
J1555.6$-$2906 & GBT, VLA & unlikely & rmb (1.7 ms)   \\
J1646.5$-$4406 & VLA      & unlikely & \omit \\
J1728.1$-$1608 & GBT, VLA & psr-C     & rm (2.5 ms) \\  % is this a binary? Might be.
J1739.3$-$2530 &  $\gamma$, GBT, VLA & psr-A     & r (1.8 s)   \\
J1803.1$+$1400 & Arecibo, VLA  & psr-C& rmb (1.5 ms)  \\
J1833.0$-$3839 & GBT, VLA & psr-C     & rmb (1.9 ms)  \\
J1843.8$-$3834 & VLA      & psr-C     & \omit \\
J2000.8$-$0300 & VLA      & unlikely & \omit \\
\hline  
\multicolumn{4}{l}{PSR Notes: g=gamma-ray pulse, r=radio pulse, m=MSP, b=binary.} 
%\multicolumn{5}{l}{PSR Notes: g=gamma-ray pulse, r=radio pulse, m=MSP, b=binary.} 
%\footnote{\url{http://tgssadr.strw.leidenuniv.nl/doku.php}}. 
\end{tabular}
\end{table}

\begin{table*}
\small
\caption{Steep-Spectrum Radio Sources Toward Fermi Unassociated Sources from the Preliminary 7-year LAT Source List}
\begin{tabular}{ccccccr}
\hline\hline
RA & DEC & $l, b$ & S$_t$ & S$_p$ & $\alpha$ & \emph{Fermi} Source \\
\hline
  06:28:44.48 ($\pm$0.13) & +05:19:17.4   ($\pm$2.0) & 205.6, $-$2.6  & 133.8$\pm$15.6 & 136.3$\pm$14.4 & $-1$.62 & J0628.6$+$0511 \\ % PGW\_0730
  12:36:46.98 ($\pm$0.16) & +11:29:03.4   ($\pm$2.2) & 289.8, 74.0     & 161.2$\pm$22.0 & 119.6$\pm$15.3 & $-$1.62 & J1236.5$+$1133 \\ %P967-012
  15:55:40.69 ($\pm$0.13) & $-$29:08:29.0 ($\pm$2.1) & 344.5, 18.5    & 133.9$\pm$14.4 & 113.0$\pm$11.7 & $-$2.5 & \textbf{J1555.6$-$2906} \\ %P86Y3595
  16:46:22.64 ($\pm$0.14) & $-$44:05:33.4 ($\pm$3.7) & 340.8, 0.8      & 198.7$\pm$26.2 & 138.3$\pm$17.3 & $-$1.86 & J1646.5$-$4406 \\ %S966-1550
%Known PSR  17:22:02.89 ($\pm$0.14) & $-$32:07:43.4 ($\pm$2.1) & 354.6, 2.5 & 384.0$\pm$40.7 & 364.1$\pm$37.2 & $-$2.25 & J1722.0$-$3204 \\ % P86Y3951
  17:27:59.08 ($\pm$0.13) & $-$16:09:09.1 ($\pm$2.1) & 8.7, 10.2        & 116.6$\pm$13.3 & 103.0$\pm$11.0 & $-$2.6 & \textbf{J1728.1$-$1608} \\ %P86Y3982
  17:39:30.70 ($\pm$0.14) & $-$25:30:17.5 ($\pm$2.2) & 2.2, 3.0          & 142.7$\pm$18.5 & 131.5$\pm$14.8 & $-$2.5 & \textbf{J1739.3$-$2530} \\ %S966-1277
  18:03:14.85 ($\pm$0.14) & +13:58:21.6   ($\pm$2.1) & 40.3,16.9        & 98.3$\pm$13.1 & 80.5$\pm$9.6      & $-$2.2 & \textbf{J1803.1$+$1400} \\ %P86Y4135
  18:33:04.45 ($\pm$0.13) & $-$38:40:46.1 ($\pm$2.1) & 356.0, $-$13.3 & 303.7$\pm$31.8 & 286.0$\pm$29.1 &$-$2.15 & \textbf{J1833.0$-$3839} \\ %P86Y4282
  18:44:35.78 ($\pm$0.13) & $-$38:40:49.6 ($\pm$2.0) & 356.9, $-$15.3 & 470.4$\pm$47.8 & 439.0$\pm$44.2 &$-$1.68 & J1843.8$-$3834 \\ %P86Y4331
  20:00:52.90 ($\pm$0.14) & $-$02:58:02.7 ($\pm$2.2) & 38.3,$-$16.8    & 99.2$\pm$13.4 & 81.6$\pm$9.8   &$-$1.68 & J2000.8$-$0300 \\ %P86Y4618
%Resolved AGN  22:57:59.21 ($\pm$0.14) & +62:33:27.3   ($\pm$2.1) & 110.3, 2.5        & 147.8$\pm$17.6 & 115.3$\pm$13.0 & $-$1.83 & J2259.1$+$6233 \\ %P86Y5242
\hline
\multicolumn{7}{p{5.3in}}{Notes: (a) Positions, flux density and peak brightness values come from the TGSS ADR1 150 MHz source catalog \citep{ijmf16}. (b) RA and DEC include a \asec{2} systematic error added in quadrature to the Gaussian fit errors. (c) The peak brightness (S$_p$) and flux density (S$_t$) values have a 10\% systematic term added in quadrature to the measurement errors. \fermi sources in boldface indicate initial pulsar detections from Table\ref{tab_candi}.}
\label{tab:steep}
\end{tabular}
\end{table*}

%\subsubsection{PGW\_0730}
\subsubsection{\textsc{Fermi} J0628.6$+$0511}

This candidate has a peak brightness of 136.3$\pm$\mjybeam{14.4} at 150 MHz. There is a moderately bright
point source at 1.4 GHz in the NVSS catalog with a flux density of 3.1$\pm$0.4 mJy (Fig.~\ref{fig:pgw}). Our follow-up VLA observations (\S\ref{obs:vla}) show an unresolved source (beam=\asec{7.9}$\times$\asec{4.9}) with a peak brightness at 3.1 GHz of 1.09$\pm$\mjybeam{0.03} and a flux density of 
1.47$\pm$0.05 mJy. A power-law fit to these data gives $\alpha=-1.62\pm{0.05}$. This VLA detection allows us to improve the position over the TGSS ADR1 value to R.A.=\hms{06}{28}{44.45} and $\delta$=\dms{05}{19}{16.8} with an uncertainty of $\pm$\asec{0.2}.

\begin{figure}
\includegraphics[width=\columnwidth]{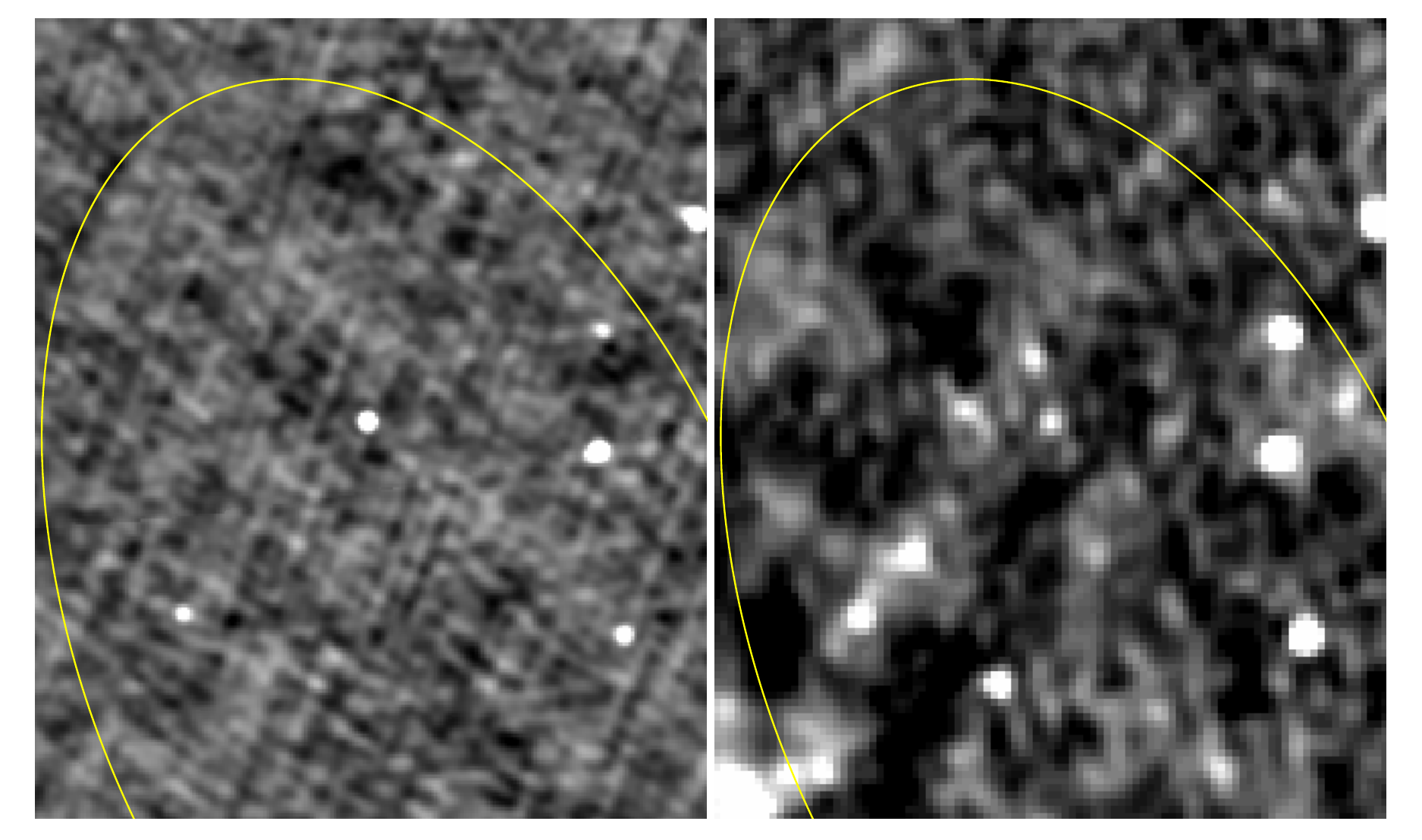}
\caption{\textsc{Fermi} J0628.6$+$0511. (Left) A steep-spectrum pulsar candidate in the center of a TGSS ADR1 image at 150 MHz. (Right) The same field from the NVSS at 1.4 GHz. The curve is the \fermi 95\% error ellipse for an unassociated gamma-ray source in this same direction, whose angular size is given in Table \ref{tab:gamma}. The field of view is \amin{19.5} by \amin{11.1}.}
\label{fig:pgw}
\end{figure}

% This source is not in the current version of the 7-year catalog. What do we do about this?
%\subsubsection{P967-012}
\subsubsection{\textsc{Fermi} J1236.5$+$1133}

The TGSS ADR1 source lies toward the largest of the \fermi error ellipses in our candidate sample.  The spectral index of $\alpha=-1.62$ is derived from the TGSS ADR1 flux density at 150 MHz and 
a moderately bright NVSS point source at 1.4 GHz with a flux density of 4.3$\pm$0.4 mJy (Fig.~\ref{fig:p967}). However, the radio source appears to be variable at 1.4 GHz since it appears in the FIRST catalog with a flux density of 8.63$\pm$0.15 mJy, double that of the NVSS value. The FIRST source appears unresolved in the catalog ($<$\asec{1.7}) and we can use the more accurate position of RA=\hms{12}{36}{47.056}, DEC=\dms{11}{29}{1.31}, with uncertainties of approximately \asec{0.15} in both coordinates. Our follow-up VLA observations (\S\ref{obs:vla}) show an unresolved source (beam=\asec{9.3}$\times$\asec{3.3}) with a peak brightness and flux density at 3.1 GHz of 2.99$\pm$\mjybeam{0.02} and 3.08$\pm$0.03 mJy, respectively. We split these S-band observations into two frequency bands at 2.69 GHz and 3.37 GHz and we measured a spectral index of  $\alpha=-1.2\pm{0.3}$, comparable to the value derived above. The TGSS position is coincident with a likely M star (R=19.15 mag) that is visible in \wise, NOMAD, SDSS and UKIDSS images. The improved VLA and FIRST positions are consistent with each other and they place the star \asec{3.5} to the northwest and hence the star is not associated with the radio source. There is no VLSSr detection. Taken together, these data suggest that the steep spectrum may be an artifact of the variability of the radio source.

\begin{figure}
\includegraphics[width=\columnwidth]{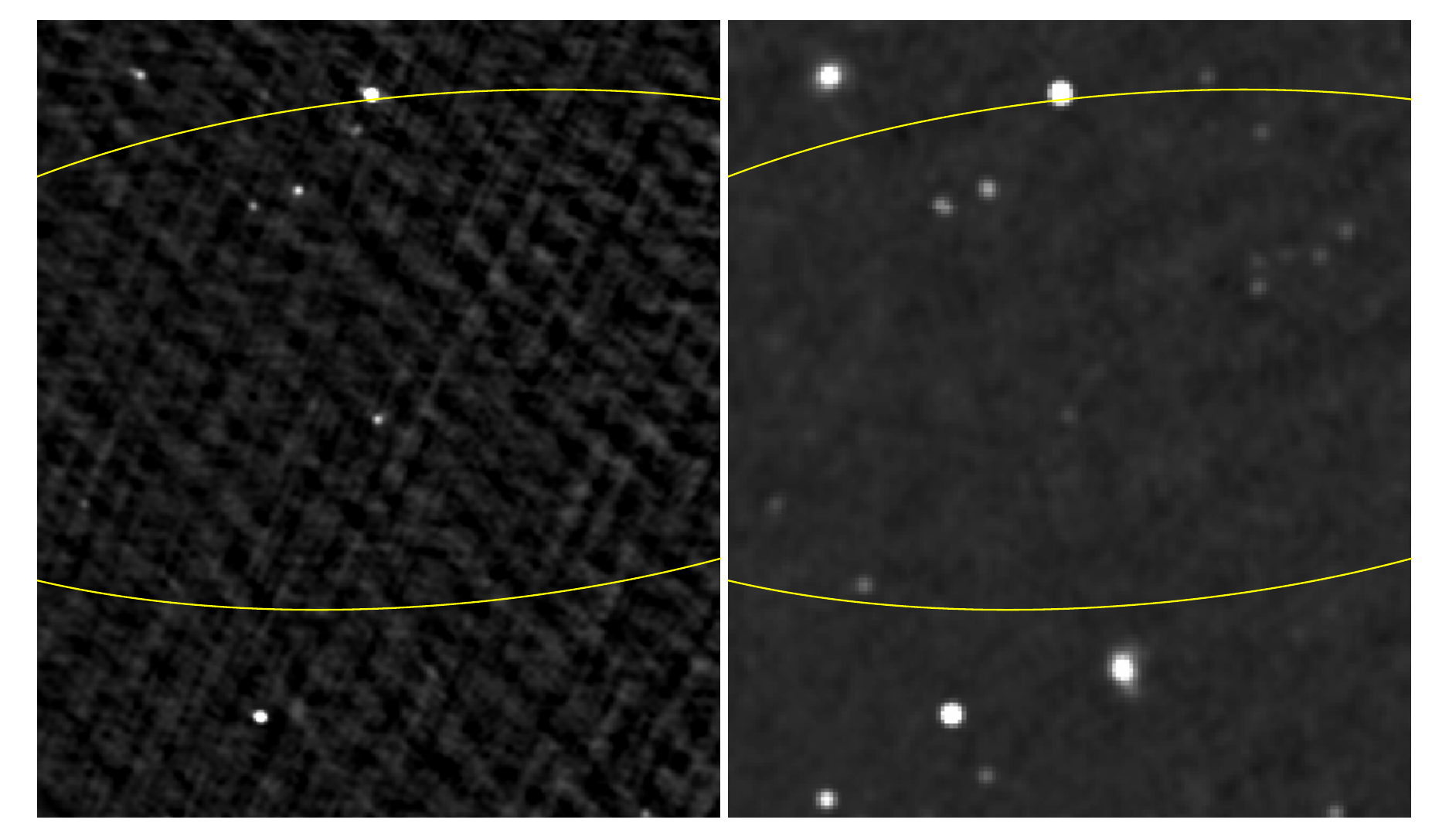}
\caption{\textsc{Fermi} J1236.5$+$1133. (Left) A steep-spectrum pulsar candidate in the center of a TGSS ADR1 image at 150 MHz. (Right) The same field from the NVSS at 1.4 GHz. The curve is the \fermi 95\% error ellipse for an unassociated gamma-ray source in this same direction, whose angular size is given in Table \ref{tab:gamma}. The field of view is \amin{23} by \amin{19.8}.}
\label{fig:p967}
\end{figure}

%\subsubsection{P86Y3595}
\subsubsection{\textsc{Fermi} J1555.6$-$2906}

There is a single TGSS ADR1 source at 150 MHz within this \fermi error ellipse but no corresponding NVSS radio source at 1.4 GHz (Fig.~\ref{fig:p86a}). The spectral index derived from the survey data and NVSS upper limit alone is  $<-2.2$. The source is included in the GLEAM source catalog \citep{hch+17}. A slightly steeper value of $\alpha=-2.5\pm 0.2$ is derived if we include the GLEAM data and the 3.1 GHz 3$\sigma$ upper limit of \mujybeam{51} from the VLA follow-up observations (\S\ref{obs:vla}).

\begin{figure}
\includegraphics[width=\columnwidth]{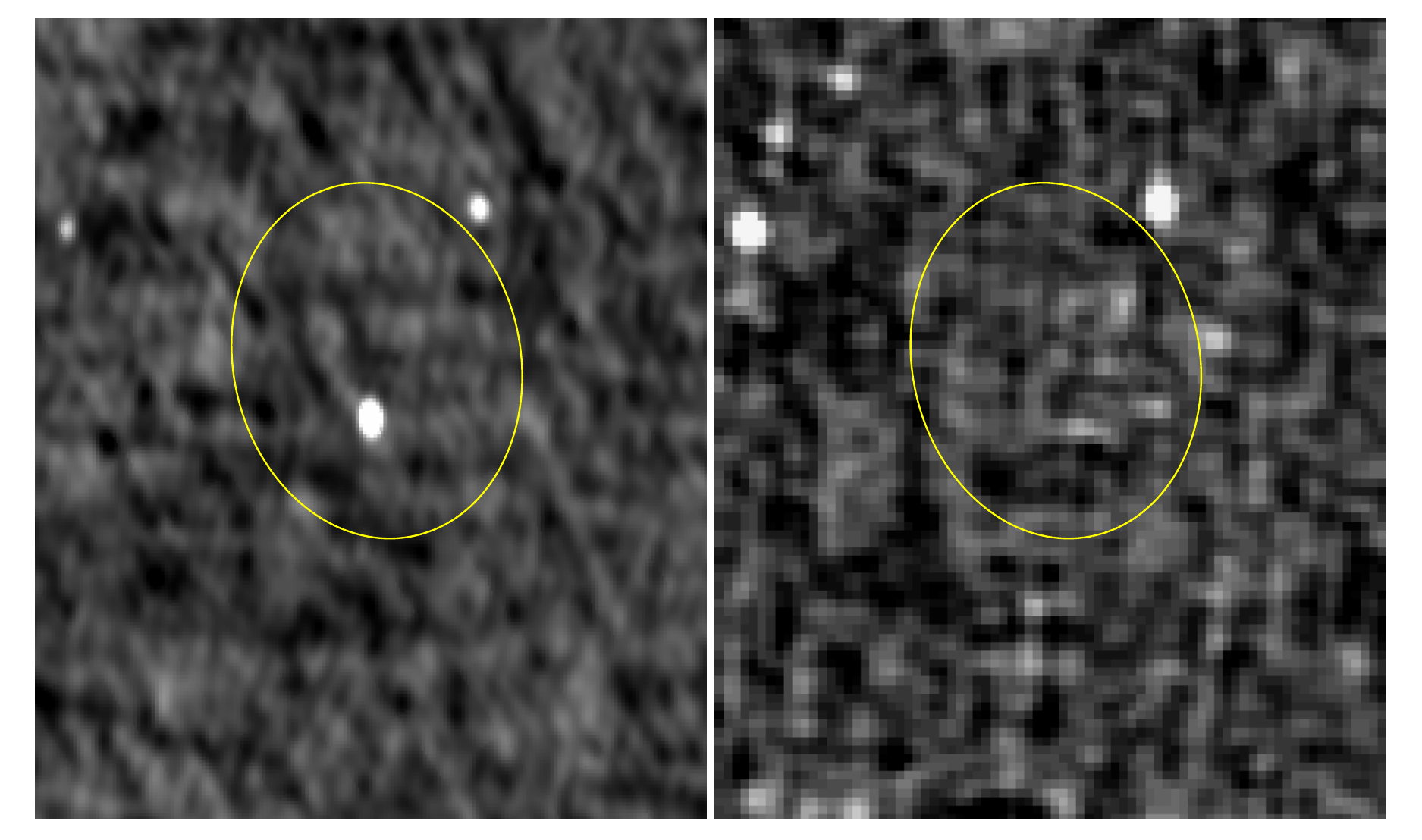}
\caption{\textsc{Fermi} J1555.6$-$2906. (Left) A steep-spectrum pulsar candidate near the center of a TGSS ADR1 image at 150 MHz.  (Right) The same field from the NVSS at 1.4 GHz. The curve indicates the \fermi 95\% error ellipse for an unassociated gamma-ray source in this same direction, whose angular size is given in Table \ref{tab:gamma}. The field of view is \amin{12.2} by \amin{10.3}.}
\label{fig:p86a}
\end{figure}

\begin{table*}
\small
\caption{Gamma-Ray Properties of Steep-Spectrum Radio Sources from the Preliminary 7-year LAT Source List}
\begin{tabular}{cccccrrr}
\hline\hline
\emph{Fermi} & Maj.$\times$Min ($\theta$) & Spec & F$_\gamma$ & Spec.&  E$_{100}$ & TS & $\Delta\theta$ \\
Source               &[deg.$\times$deg. ($\circ$)] & Type          & \omit & Index & [erg cm$^{-2}$ s$^{-1}$] & \omit & [deg.] \\
\hline
J0628.6$+$0511  & 0.34$\times$0.20 (22.0) & PL & 2.42E-13 & 2.59 & 6.95E-12    &  27.9 & 0.13    \\ %PGW\_0730
J1236.5$+$1133    & 0.63$\times$0.25 (-85.1) & PL & 7.15E-13 & 2.64 & 2.44E-12 &  29.7 & 0.09  \\ %P967-012
J1555.6$-$2906    & 0.09$\times$0.07 (11.7)  & PL & 5.94E-13 & 2.51 & 5.67E-12 &  87.7 & 0.03  \\  %P86Y3595
J1646.5$-$4406    & 0.05$\times$0.05 (74.2)  & PL & 1.04E-13 & 2.49 & 1.41E-11 &  31.6 & 0.04  \\ %S966-1550
%PSR J1722.0$-$3204   & 0.08$\times$0.08 (47.5)  & PL & 1.19E-13 & 2.43 & 1.016E-11 & \omit & 0.06 \\ %P86Y3951
J1728.1$-$1608    & 0.08$\times$0.07 (88.9)  & LP & 5.65E-13 & 1.87 & 3.27E-12 &  66.7 & 0.05  \\ %P86Y3982
J1739.3$-$2530    & 0.06$\times$0.05 (55.0)  & LP & 1.02E-12 & 2.38 & 1.01E-11 & 105.4 & 0.04  \\ %S966-1277
J1803.1$+$1400    & 0.11$\times$0.09  (-32.3) & PL & 5.74E-14 & 2.18 & 2.77E-12&  30.9 &  0.04 \\ %P86Y4135
J1833.0$-$3839    & 0.11$\times$0.08  (63.3) & PL & 1.33E-13 & 2.36 & 2.90E-12 &  31.0 & 0.02  \\ %P86Y4282
J1843.8$-$3834   & 0.3 $\times$0.26  (10.0)  & PL & 2.59E-13 & 2.55 & 2.20E-12 &  13.1 & 0.19  \\ %P86Y4331
 J2000.8$-$0300   & 0.41$\times$0.22  (12.6) & PL & 4.75E-13 & 2.74 & 3.45E-12 &  18.5 & 0.04  \\ %P86Y4618
%AGN J2259.1$+$6233  & 0.17$\times$0.10 (80.7)  & PL & 1.63E-13 & 2.65 & 7.76E-12 & \omit & 0.13 \\ %P86Y5242
\hline
\multicolumn{8}{p{4.9in}}{Notes: (a) The gamma-ray flux density (100 MeV to 300 GeV) F$_\gamma$ is in units of ph. cm$^{-2}$ MeV$^{-1}$ s$^{-1}$. (b) Spectral Type is either power-law (PL) or a log-parabola (LP).}
\label{tab:gamma}
\end{tabular}
\end{table*}

%\subsubsection{S966-1550}
\subsubsection{\textsc{Fermi} J1646.5$-$4406}

This candidate lies \amin{16} to the southeast from the center of the shell-like Galactic supernova remnant G\,341.2+0.9 that is claimed to be associated with PSR\,B1643$-$43 and its pulsar wind nebula \citep{fgw94,gfgv01}. The TGSS ADR1 candidate is embedded  within a ``spur" of emission that begins at the southeastern edge of SNR G\,341.2+0.9 and runs north-south (Fig.~\ref{fig:s966a}). At this position in the SUMSS image at 843 MHz  there is a point source with a peak brightness of \mjybeam{10}. The source is visible at 330 MHz but the resolution of the synthesized beam in \citet{fgw94} is too poor to measure the flux density of the point source separate from the extended emission. Our follow-up \atca observations at 2.1 GHz (\S\ref{obs:atca}) required two epochs taken at different elevation angles because the side-lobes of bright, nearby sources made flux density measurements difficult. In the combined image we detect a source with a flux density of 1.4$\pm${0.2} mJy. A power-law fit to the three detections gives $\alpha=-1.86\pm{0.12}$. The ATCA detection allows for an improved position of R.A.=\hms{16}{46}{22.77} and $\delta$=\dms{$-$44}{05}{40.8} with an uncertainty of $\pm$\asec{0.5}. The TGSS and ATCA declination values differ by about 2$\sigma$, or about one ATCA synthesized beam. 

We note that the ratio of the flux density to peak brightness at 150 MHz in Table \ref{tab:steep} is 1.4. Such high ratios would normally suggest that the radio source is marginally resolved \citep[see \S{4.6} of][]{ijmf16} and thus is not a viable pulsar candidate. However, as noted in \S\ref{method}, some caution is warranted when searching for pulsars as strong scintillation could modulate the flux density during the observations, making such ratios unreliable and in some cases creating radial image artifacts \citep{fjmi16} similar to those visible north of the point source in the 150 MHz image of Fig.\ref{fig:s966a}.

\begin{figure}
\includegraphics[width=\columnwidth]{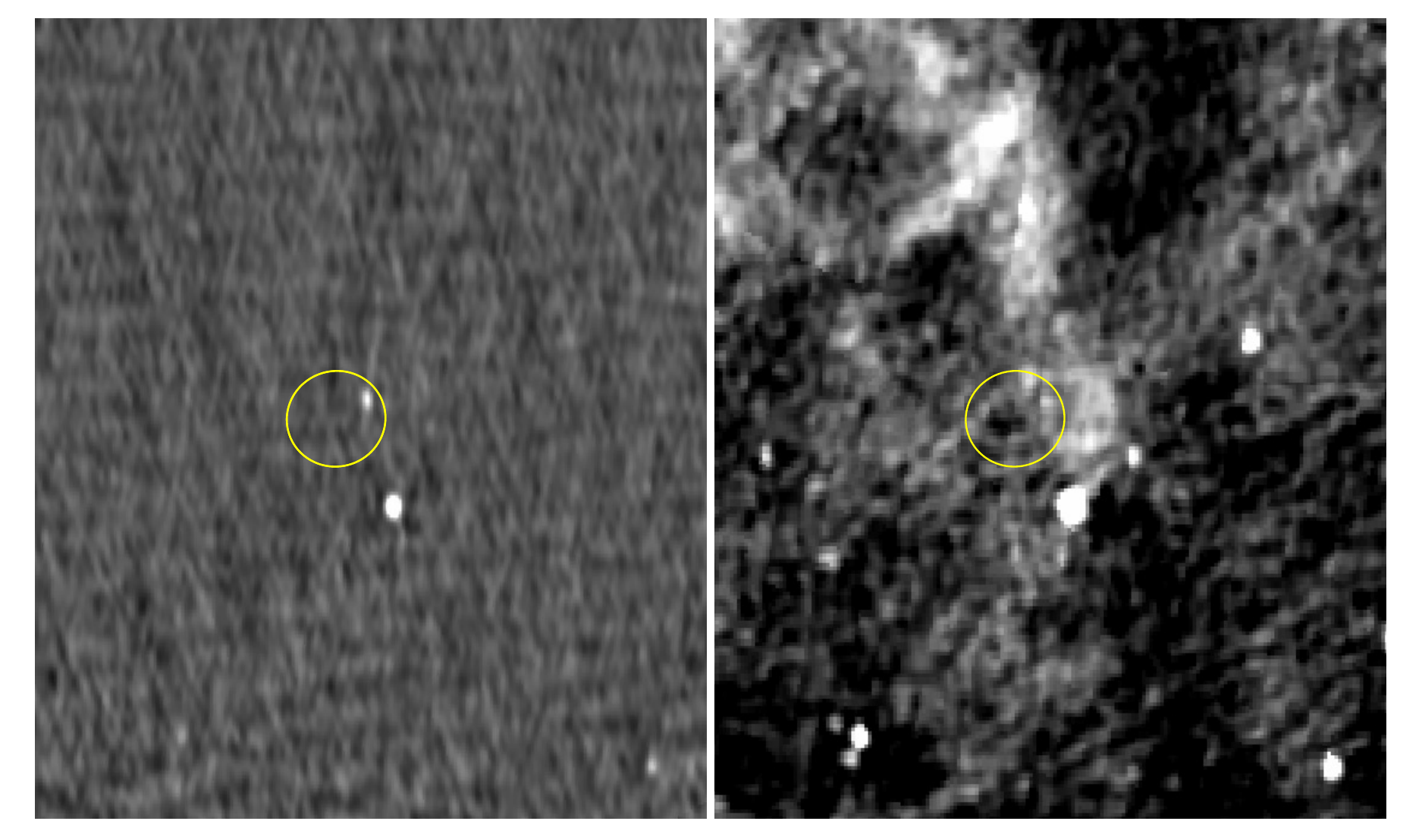}
\caption{\textsc{Fermi} J1646.5$-$4406. (Left) A steep-spectrum pulsar candidate near the center of a TGSS ADR1 image at 150 MHz. (Right) The same field from the SUMSS at 843 MHz. The shell-like Galactic supernova remnant G\,341.2+0.9 
%and its pulsar  PSR\,B1643$-$43 are in
is in the upper left corner of this image. The curve indicates the \fermi 95\% error ellipse for an unassociated gamma-ray source in this same direction, whose angular size is given in Table \ref{tab:gamma}. The field of view is \amin{24.6} by \amin{20.7}.}
\label{fig:s966a}
\end{figure}

%\subsubsection{P86Y3951}
%\subsubsection{\textsc{Fermi} J1722.0$-$3204}

%P86Y3951: The steep-spectrum candidate toward this \fermi unassociated source is a known pulsar PSR\,B1718$-$32. This is an old, slow pulsar with a period of 477 ms with a spin-down energy of 2.3$\times 10^{32}$ erg s$^{-1}$. Given the low spin-down energy, this pulsar likely not associated and is likely a line-of-sight coincidence with the gamma-ray source.

%\begin{figure}
%\includegraphics[width=\columnwidth]{p86y3951_tgss_sumss.eps}
%\caption{P86Y3951. (Left) TGSS. (Right) SUMSS.}
%\label{fig:p86b}
%\end{figure}

%\subsubsection{P86Y3982}
\subsubsection{\textsc{Fermi} J1728.1$-$1608}

There is an unresolved TGSS ADR1 radio source at 150 MHz within this small \fermi unassociated error ellipse, but only an upper limit on any emission in the NVSS images at 1.4 GHz (Fig.~\ref{fig:p86c}). There is no corresponding point source in the VLSSr image at 74 MHz at this location but there is some extended emission \amin{1.5} to the northeast that has no counterpart in the NVSS or TGSS ADR1 images. Our follow-up VLA observations (\S\ref{obs:vla})  provide a 3$\sigma$ upper limit at 3.1 GHz of \mujybeam{37}. The upper limit on the spectral index implied by these limits is $\alpha<-2.6$.

\begin{figure}
\includegraphics[width=\columnwidth]{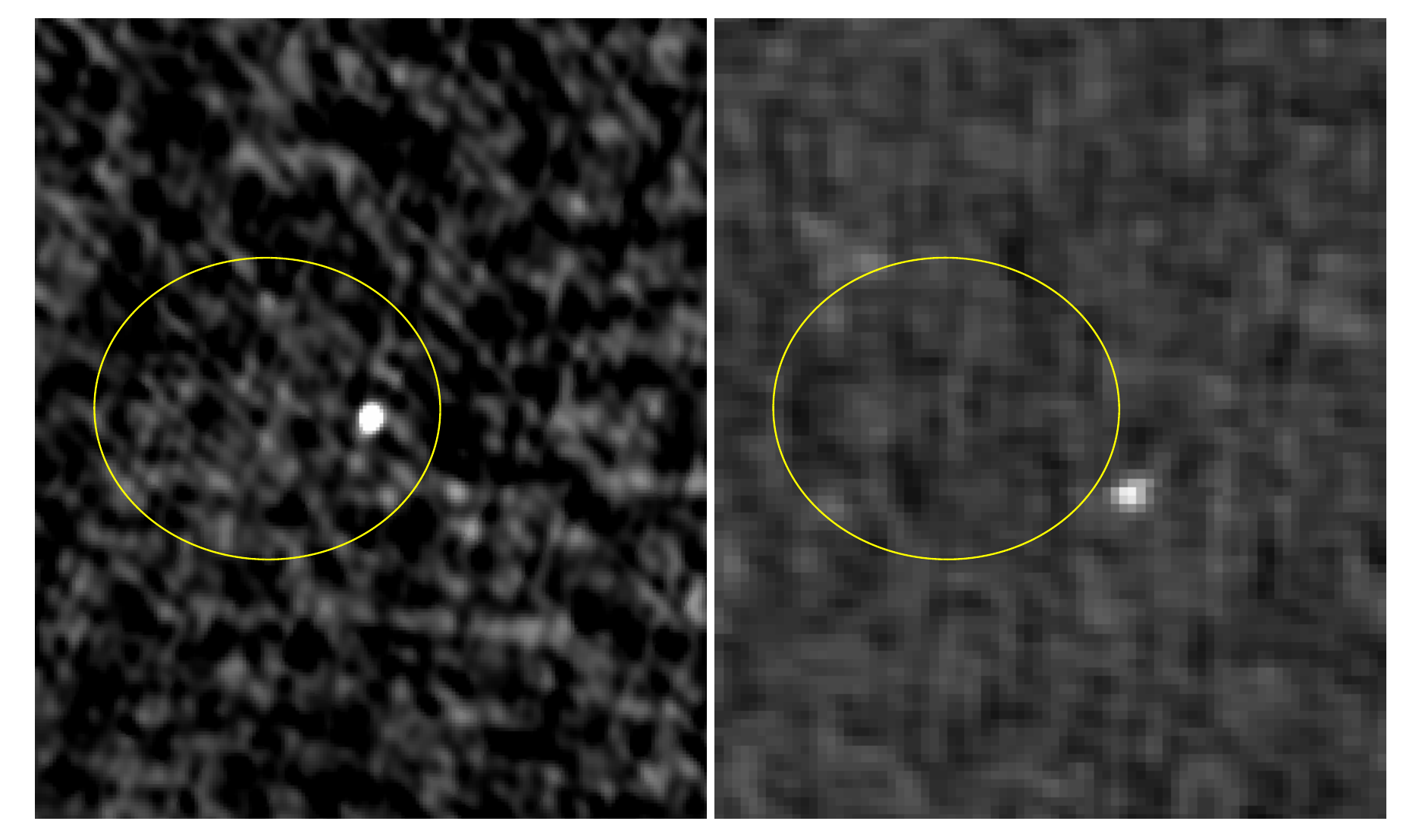}
\caption{\textsc{Fermi} J1728.1$-$1608. (Left) A steep-spectrum pulsar candidate near the center of a TGSS ADR1 image at 150 MHz.  (Right) The same field from the NVSS at 1.4 GHz. The curve indicates the \fermi 95\% error ellipse for an unassociated gamma-ray source in this same direction, whose angular size is given in Table \ref{tab:gamma}. The field of view is \amin{11.4} by \amin{9.6}.}
\label{fig:p86c}
\end{figure}

%\subsubsection{S966-1277}
\subsubsection{\textsc{Fermi} J1739.3$-$2530}

There is an unresolved TGSS ADR1 radio source with a peak brightness at 150 MHz of 131.5$\pm$\mjybeam{14.8} within this small \fermi unassociated error ellipse, but only an upper limit on any emission in NVSS images at 1.4 GHz of $<$\mjybeam{0.3} (Fig.~\ref{fig:s966b}). Our follow-up VLA observations (\S\ref{obs:vla})  provide only a 3$\sigma$ upper limit at 3.1 GHz of \mujybeam{70}. The upper limit on the spectral index implied by these limits is $\alpha<-2.5$.

\begin{figure}
\includegraphics[width=\columnwidth]{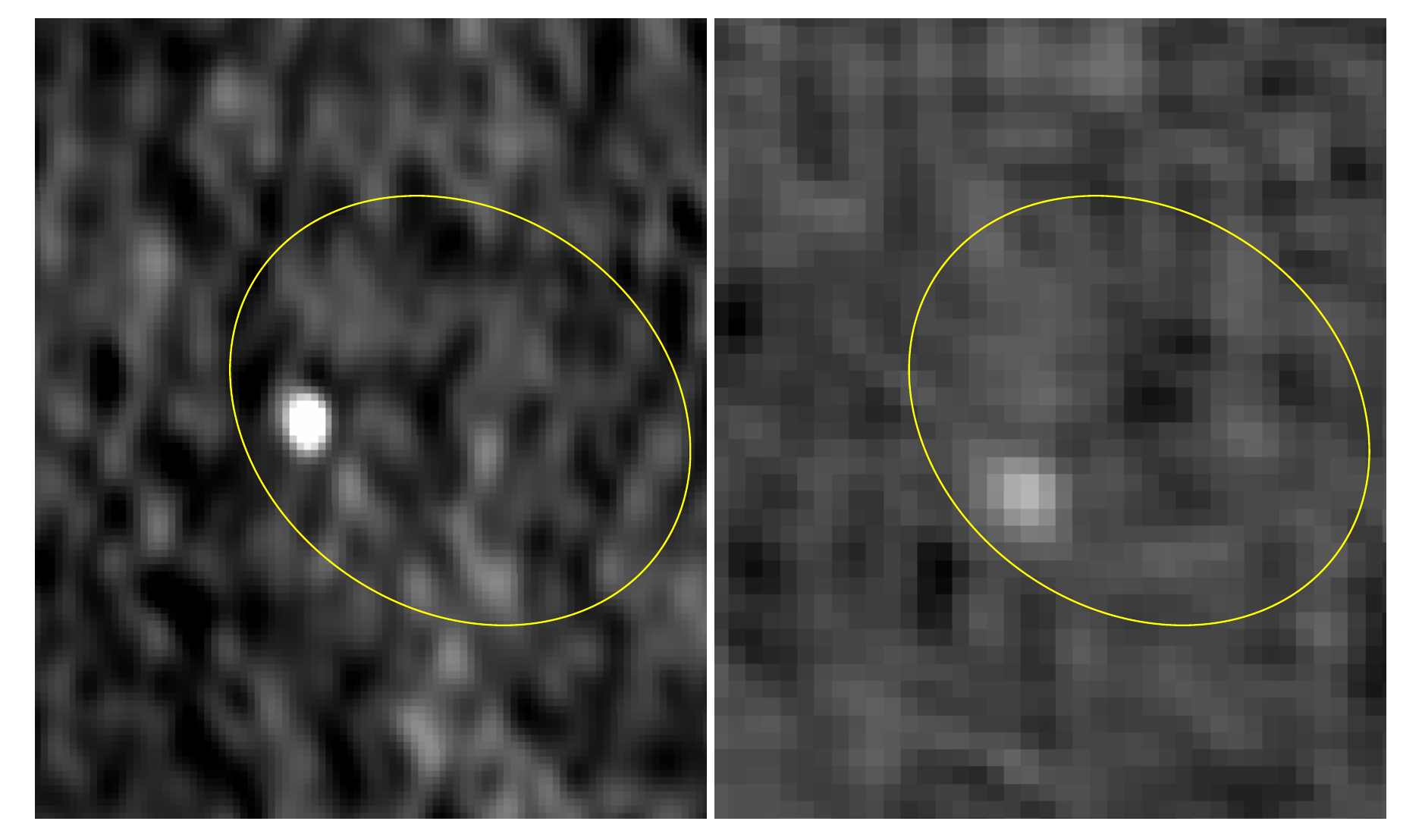}
\caption{\textsc{Fermi} J1739.3$-$2530. (Left) A steep-spectrum pulsar candidate left of center in a TGSS ADR1 image at 150 MHz.  (Right) The same field from the NVSS at 1.4 GHz. The curve indicates the \fermi 95\% error ellipse for an unassociated gamma-ray source in this same direction, whose angular size is given in Table \ref{tab:gamma}. The field of view is \amin{5.8} by \amin{4.9}.}
\label{fig:s966b}
\end{figure}

%\subsubsection{P86Y4135}
\subsubsection{\textsc{Fermi} J1803.1$+$1400}

There is an unresolved TGSS ADR1 radio source at 150 MHz within this small \fermi unassociated error ellipse, but only an upper limit on any emission in NVSS images at 1.4 GHz of $<$\mjybeam{0.3} (Fig.~\ref{fig:p86d}). At this same position on the VLSSr image at 74 MHz there is a marginal detection with a peak brightness of 275$\pm$150 mJy beam$^{-1}$. Our follow-up VLA observations (\S\ref{obs:vla})  provide only a 3$\sigma$ upper limit at 3.1 GHz of \mujybeam{72}. The upper limit on the spectral index implied by these limits is $\alpha<-2.2$.

\begin{figure}
\includegraphics[width=\columnwidth]{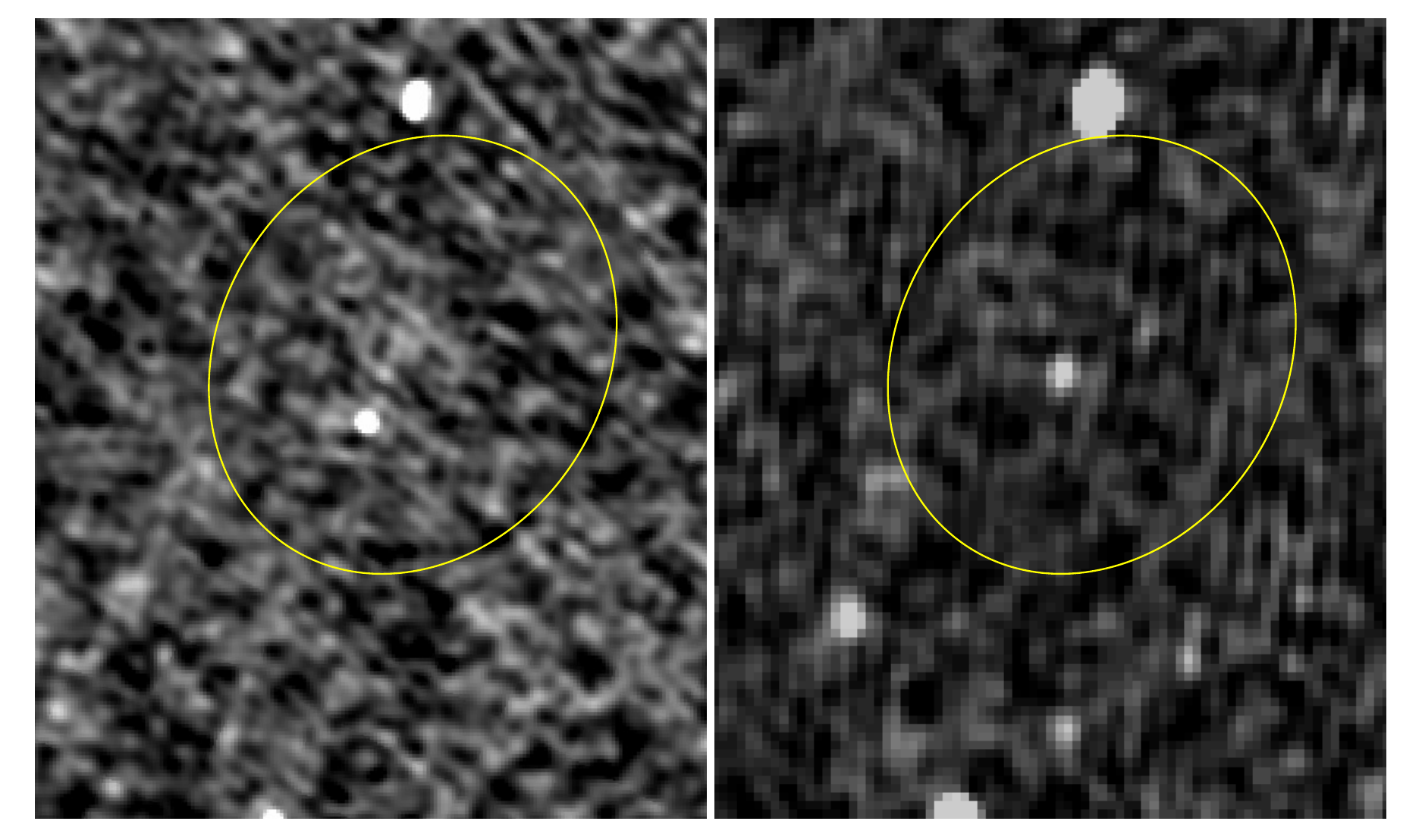}
\caption{\textsc{Fermi} J1803.1$+$1400. (Left) A steep-spectrum pulsar candidate at the center of a TGSS ADR1 image at 150 MHz.  (Right) The same field from the NVSS at 1.4 GHz. The curve indicates the \fermi 95\% error ellipse for an unassociated gamma-ray source in this same direction, whose angular size is given in Table \ref{tab:gamma}.The field of view is \amin{11.4} by \amin{9.7}.}
\label{fig:p86d}
\end{figure}
  
%\subsubsection{P86Y4282}
\subsubsection{\textsc{Fermi} J1833.0$-$3839}

There is a bright TGSS ADR1 source at 150 MHz (S$_t$=303.7$\pm$31.8 mJy) near the center of the \fermi error ellipse but there is no corresponding radio source at this location in the SUMSS image at 843 MHz or the NVSS image at 1.4 GHz (Fig.~\ref{fig:p86dd}). A conservative two-point $\alpha$ is calculated in Table \ref{tab:steep} adopting a 5$\sigma$ limit of 2.5 mJy at 1.4 GHz.  Our follow-up \atca observations at 2.1 GHz (\S\ref{obs:atca}), giving an upper limit of \mjybeam{0.75}, suggests a slightly steeper spectral index. A mosaic joint is visible along a diagonal in the TGSS ADR1 image in Fig. \ref{fig:p86dd}, a few arcminutes north of the candidate radio source, but it is at the level of the rms noise and appears to have no effect on the flux density measurement. 
% I got the rms limit from the images that Paul provided. rms = 250 uJy and beam=11.5"x3.8" (-22deg)

\begin{figure}
\includegraphics[width=\columnwidth]{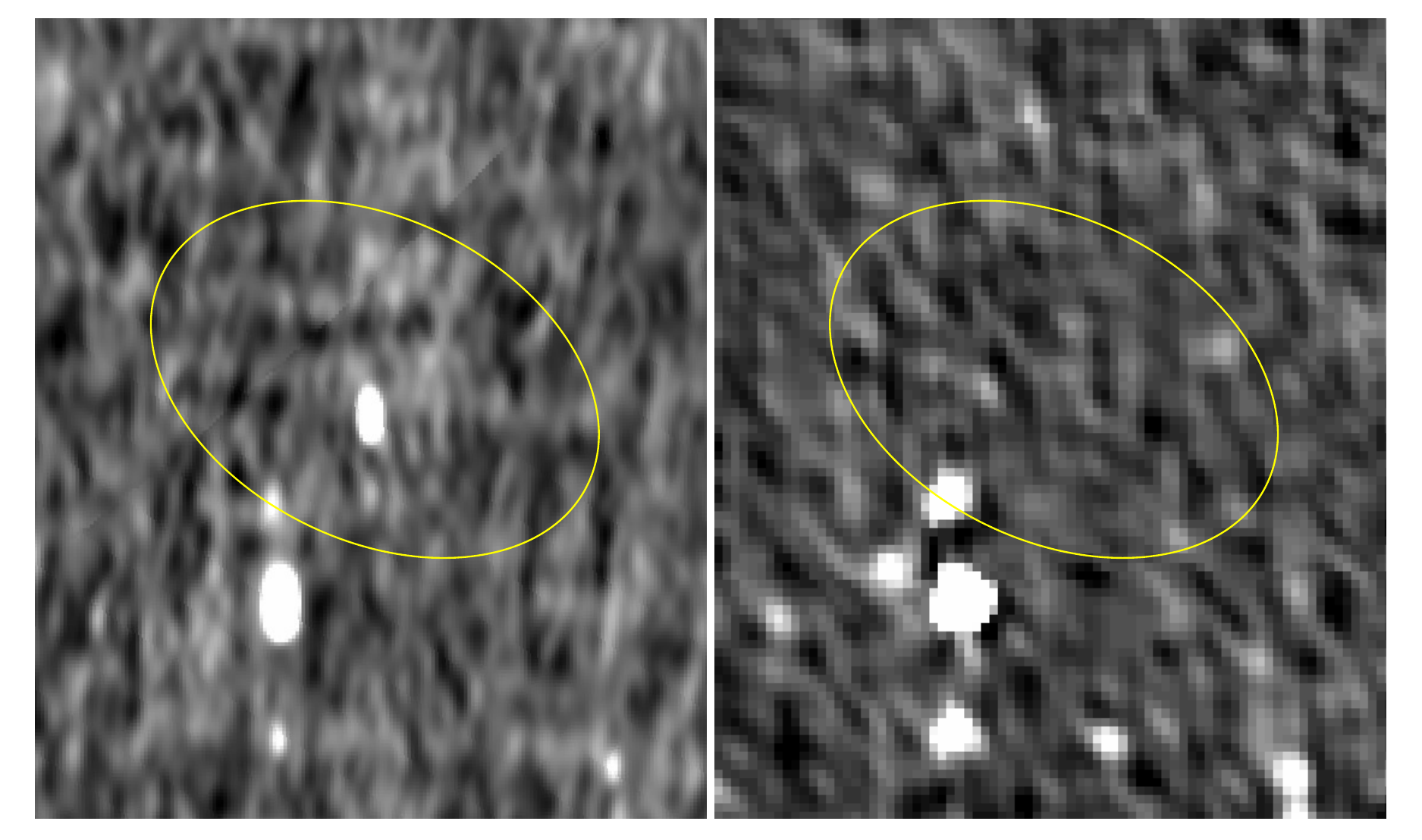}
\caption{\textsc{Fermi} J1833.0$-$3839. (Left)  A steep-spectrum pulsar candidate near the center of a TGSS ADR1 image at 150 MHz.  (Right) The same field from the NVSS at 1.4 GHz. The curve indicates the \fermi 95\% error ellipse for an unassociated gamma-ray source in this same direction, whose angular size is given in Table \ref{tab:gamma}. The field of view is \amin{11.2} by \amin{9.5}.}
\label{fig:p86dd}
\end{figure}

%\subsubsection{P86Y4331}
\subsubsection{\textsc{Fermi} J1843.8$-$3834}

% I refit the position using Paul's fits images on dropbox. Beam is 13.7x3.7" (-29deg). The error i slarger than the original 0.15" and it makes better sense I think, given the beam size and S/N.
This candidate in Fig.~\ref{fig:p86e} is detected strongly in the TGSS ADR1 (S$_t$=470.4$\pm$47.8 mJy) and also in SUMSS at 843 MHz (S=21.1$\pm$1.2 mJy) and NVSS at 1.4 GHz (S=8.6$\pm$0.5 mJy). Our follow-up \atca observations (\S\ref{obs:atca}) show an unresolved source (beam=\asec{13.7}$\times$\asec{3.7}) with a peak brightness at 2.1 GHz of 4.14$\pm$\mjybeam{0.13}. Based on these four flux densities we recompute the spectral index to be $\alpha=-1.79\pm0.05$. The \atca detection leads to an improved position of R.A.=\hms{18}{44}{35.77} and $\delta$=\dms{$-$38}{40}{50.20} with an uncertainty of $\pm$\asec{1.2} in declination and about a factor of two better in right ascension.  The source is included in the GLEAM source catalog. The spectral index derived from the survey data alone is $-1.4\pm 0.4$ \citep{hch+17}. If we fit to all the data, including GLEAM (Fig.~\ref{fig:spectra}), the spectral index $\alpha=-1.68\pm 0.10$.

% I added 3% flux density uncertaity in quadrature to get the true uncertainty in the ATCA flux density value.

\begin{figure}
\includegraphics[width=\columnwidth]{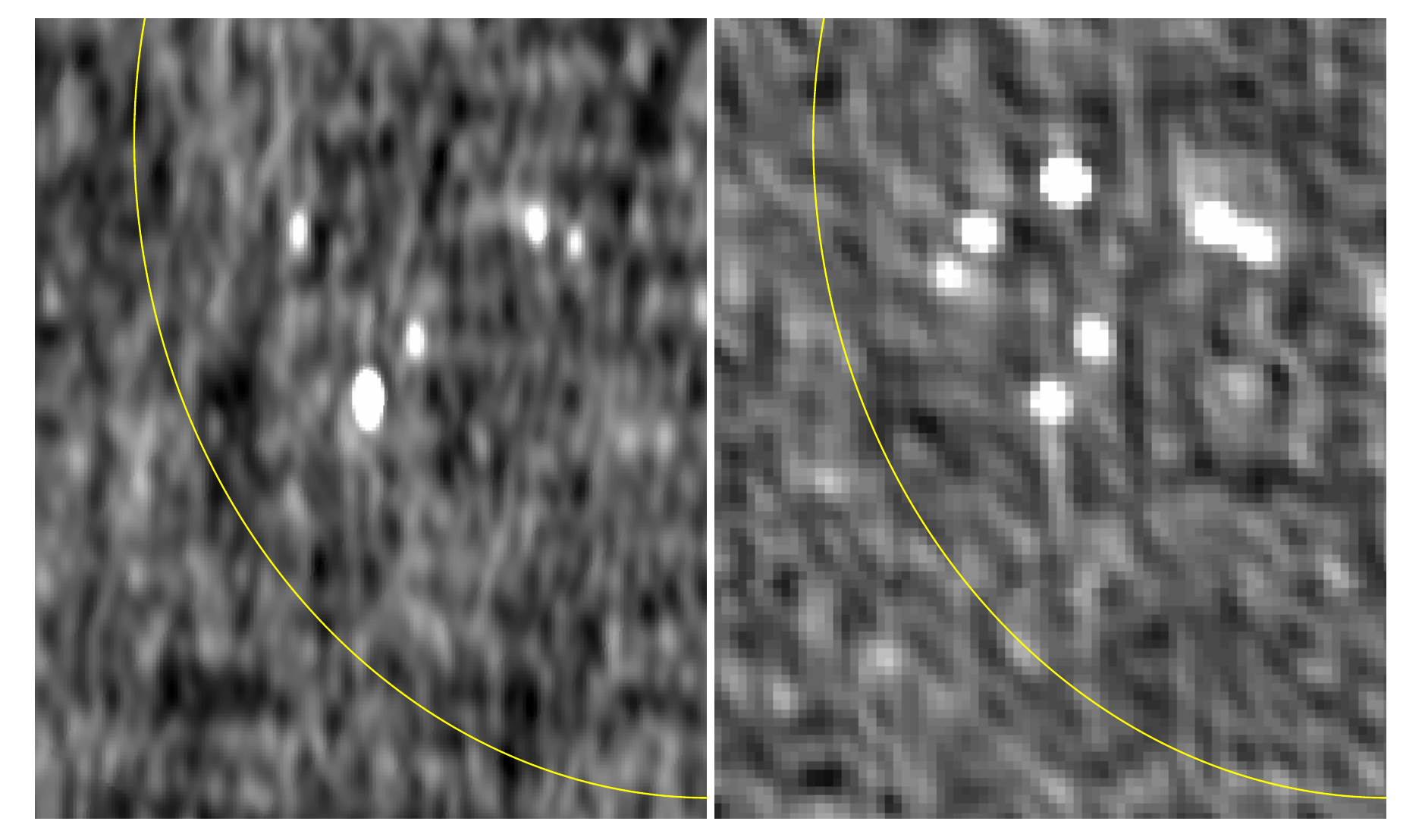}
\caption{\textsc{Fermi} J1843.8$-$3834. (Left) A steep-spectrum pulsar candidate near the center of a TGSS ADR1 image at 150 MHz.  (Right) The same field from the NVSS at 1.4 GHz. The curve indicates the \fermi 95\% error ellipse for an unassociated gamma-ray source in this same direction, whose angular size is given in Table \ref{tab:gamma}. The field of view is \amin{21.9} by \amin{18.5}.}
\label{fig:p86e}
\end{figure}

\begin{figure}
\includegraphics[width=\columnwidth]{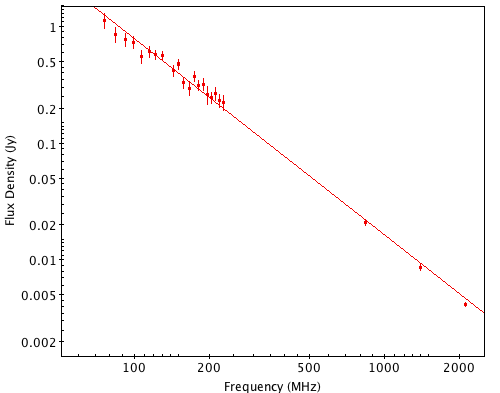}
\caption{
A radio spectrum for pulsar candidate \textsc{Fermi}\,J1843.8$-$3834. In addition to TGSS ADR1 flux densities, we include GLEAM, SUMSS, NVSS and ATCA (follow-up) observations. The source is unresolved at arcsecond resolution and has a power-law spectral slope of $\alpha=-1.68\pm{0.10}$.}
\label{fig:spectra}
\end{figure}

%\subsubsection{P86Y4618}
\subsubsection{\textsc{Fermi} J2000.8$-$0300}

This candidate lies near the center of a moderately large \fermi error circle (Fig.~\ref{fig:p86f}). At 150 MHz its flux density is 99.2$\pm$13.4 mJy. There is a faint NVSS source at 1.4 GHz with a flux density of 2.3$\pm$0.5 mJy. The source is not visible in the VLSSr at 74 MHz with a peak brightness  of $<$\mjybeam{315}.  Our follow-up VLA observations (\S\ref{obs:vla}) show an unresolved source (beam=\asec{6.6}$\times$\asec{3.9}) with a peak brightness at 3.1 GHz of 660$\pm$\mujybeam{17} and a flux density of 0.75$\pm$0.03 mJy. A power-law fit to these data gives $\alpha=-1.62\pm{0.05}$. This VLA detection allows us to improve the position over the TGSS ADR1 value to R.A.=\hms{20}{00}{52.75} and $\delta$=\dms{$-$02}{58}{3.5} with an uncertainty of $\pm$\asec{0.2}.  There is a weak source in the GLEAM source catalog at this position. The spectral index derived from the survey data alone is $-1.5\pm 1.7$ \citep{hch+17}. If we fit to all the available data, including GLEAM,  the spectral index $\alpha=-1.68\pm 0.10$.

\begin{figure}
\includegraphics[width=\columnwidth]{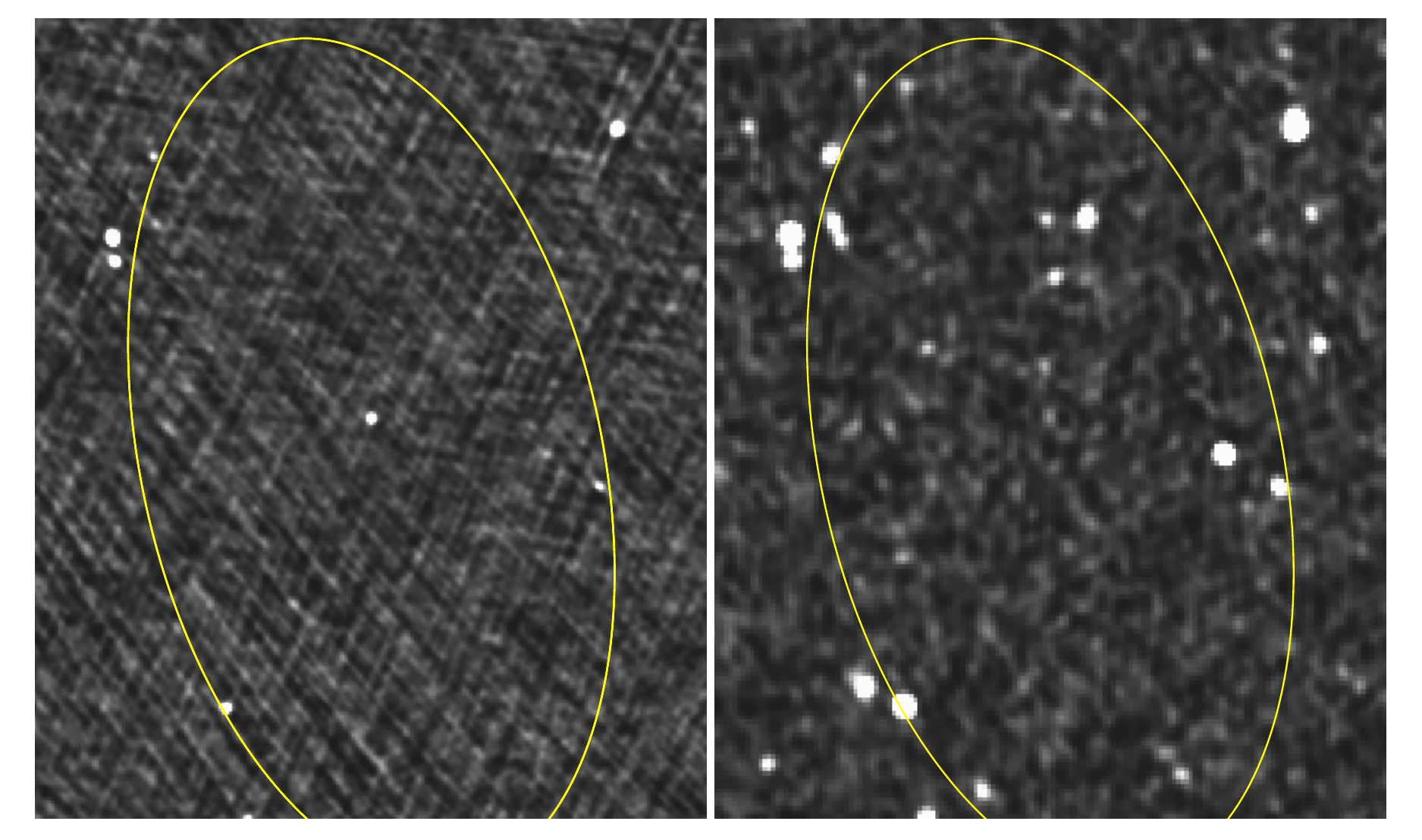}
\caption{\textsc{Fermi} J2000.8$-$0300. (Left) A steep-spectrum pulsar candidate near the center of a TGSS ADR1 image at 150 MHz.  (Right) The same field from the NVSS at 1.4 GHz. The curve indicates the \fermi 95\% error ellipse for an unassociated gamma-ray source in this same direction, whose angular size is given in Table \ref{tab:gamma}. The field of view is \amin{23.6} by \amin{19.9}.}

\label{fig:p86f}
\end{figure}
  
%\subsubsection{P86Y5242} 
%\subsubsection{J2259.1$+$6233}

%This candidate lies near the edge of a moderately large \fermi error circle in the Cepheus B star forming region with S$_t$=147.8$\pm$17.6 mJy. Point-like emission is visible at this location at both 327 MHz in the WENSS image and at 1.4 GHz in the NVSS with peak brightness of 91$\pm$7 mJy beam$^{-1}$ and 3.1 $\pm$0.5 mJy beam$^{-1}$, respectively.  There is no source at 74 MHz in the VLSSr image with a limit of $<$\mjybeam{300}. There is a faint second source that is \amin{2} to the southwest with an inverted spectral index. It reaches a peak brightness of \mjybeam{78} at 1.4 GHz.

%\begin{figure}
%\includegraphics[width=\columnwidth]{p86y5242_tgss_nvss.eps}
%\caption{P86Y5242. (Left) A steep-spectrum pulsar candidate near the center of a TGSS ADR1 image at 150 MHz.  (Right) The same field from the NVSS at 1.4 GHz. The steep-spectrum candidate is bright at 150 MHz but faint in the 1.4 GHz image. The bright source in the NVSS image at 1.4 GHz is an inverted spectrum radio source \amin{2} to the southwest of the steep-spectrum candidate. The curve indicates the \fermi 95\% error ellipse for an unassociated gamma-ray source in this same direction, whose angular size is given in Table \ref{tab:gamma}.}
%\label{fig:p86g}
%\end{figure}

\section{Pulsation Searches}\label{pulse}

After the creation of our list of unresolved steep-spectrum candidates (Tables \ref{tab_candi} and \ref{tab:steep}), we encouraged pulsar search teams, including members of the \textit{Fermi} Pulsar Search Consortium (PSC) to search them for radio and gamma-ray pulsations. The PSC has been doing this with great success using target lists consisting of \fermi LAT unassociated sources with spectra that are similar to the spectra of typical gamma-ray pulsars \citep{hrm+11,kcj+12,rrc+13,ckr+15,cck+16}. With our new technique, these searches could be expanded to LAT sources with poor localizations or poorly determined spectra that would not otherwise be highly ranked targets (see \S\ref{discuss}). For the gamma-ray pulsation searches \citep{cwp+16}, the precise interferometric positions greatly reduce the parameter space that needs to be covered, significantly improving the sensitivity of the searches. In some cases new observations were made and in others PSC members had already targeted the \fermi sources associated with our candidates, though we did not know the pulsar nature of any of our sources in advance.

These searches have, so far, resulted in the discovery of 6 millisecond pulsars and one slow pulsar, as noted in Table \ref{tab_candi}. The details of those discoveries will be published elsewhere (Clark et al., Ransom et al., Deneva et al., Camilo et al. in prep)\footnote{See also the WVU Galactic Millisecond Pulsar List (\url{http://astro.phys.wvu.edu/GalacticMSPs/}) and the  Public List of LAT Detected Pulsars (\url{
https://confluence.slac.stanford.edu/display/GLAMCOG/Public+List+of+LAT-Detected+Gamma-Ray+Pulsars}).}, as follow up is required to characterize their spin and orbital parameters, and confirm (or not) their identification with the target LAT sources. We include them here as a demonstration of the efficiency of this method for finding pulsars. We note that the data taking and analysis of these searches are not yet complete and more discoveries among our candidate list are likely.

We also observed with a new VLA pulsar observing system, an emerging capability on the VLA that has not been widely used for pulsar searches (Project code 16B-434). The array was in a Move configuration between A and D on the 4th and 6th of February 2017. The VLA was operated as a phased array by observing a nearby point source calibrator and tying the wavefront phase corrections to a common antenna near the center of the array. Phased array data were collected using the real-time pulsar processing backend (YUPPI). We observed in the frequency range of 1-2 GHz (L band) where YUPPI produces 
%fast-dump 
spectra with 1 GHz total bandwidth (32 x 32-MHz subbands) with 8-bit samplers in VLBI data interchange format (VDIF). The integration time on each pulsar was typically about 30 min. Pulsar signals were searched using standard pulsar software.

These pulsation searches are summarized in Table \ref{tab_candi}. In column two we list whether a gamma-ray pulsation search was carried out ($\gamma$) and the names of the telescopes involved in the radio searches. When more than one radio telescope is listed, the telescope that made the discovery is first. A blank entry means no pulsation search has been carried out to date. Column three is a short-hand rating of each \fermi source based on how likely it is to be a pulsar, based on its gamma-ray spectral properties \citep{ckr+15}.  In this scheme, likely pulsars are listed in order of decreasing quality, i.e. psr-A (high), psr-B (good), psr-C (ambiguous), psr-D (poor, but cannot be ruled out) and "Unlikely" to be a pulsar. Detections are given in column four. Initial properties (i.e. gamma-ray pulsations=g, radio pulsations=r, millisecond pulsar=m, and binarity=b) are given along with the pulsar rotation period in parenthesis. 

In brief, of the six MSPs, one (PSR\,J1833$-$3840) had been discovered in 2015 but is yet unpublished (F. Camilo, priv. comm.). All of the radio pulsars that were discovered were made by the Arecibo Telescope and the Green Bank Telescope (GBT). The VLA confirmed pulsations toward  J1555.6$-$2906 but made no new discoveries on its own. Two MSPs have gamma-ray pulsations, making them firm \fermi identifications \citep{c+18}. There are at least three binaries among the MSPs. The initial properties of the remaining MSPs suggest that they too are associated with the \fermi emission but long-term pulse timing is needed to confirm. There is one newly discovered slow, normal pulsar (PSR\,J1739$-$2530) that appears to be a nulling pulsar whose slow period suggests that it is likely a line-of-sight coincidence and not a gamma-ray source.

\section{Discussion and Conclusions}\label{discuss}

We have described an image-based method for identifying pulsar candidates in radio sky surveys using only two selection criteria: compactness and spectral slope. With follow-up arcsecond imaging we were able to eliminate false positives and define a sample of 16 promising candidates from the 3FGL catalog and a 7-year \fermi LAT source list. While these compact, steep-spectrum radio sources within the error ellipses of the \fermi LAT unassociated sources are strong candidates for pulsars, it does not follow that they are gamma-ray-emitting pulsars. An identical search of the same total area but in random directions on the sky, would result in a similar number of candidates. Establishing a firm gamma-ray association requires pulsation searches.

Preliminary pulsation searches of these candidates have found six MSPs and one normal rotation-powered pulsar. Two of the MSPs have confirmed gamma-ray pulsations. The remaining MSPs require further timing before any associations can be definitively claimed. Adding in the recent MSP found toward the \fermi GeV excess near the Galactic center \citep{bdf+17}, the pulsar detection efficiency using this technique is 40\%. Based on just the initial finding of four MSPs from the 7-year \fermi LAT source list, this gives a lower limit of 22 MSPs per 1000 square degrees. The success rate exceeds blind surveys with their yield of approximately 1 MSP per 1000 square degree \citep{slr+14,bcm+16}, and it compares favorably with recent pulsation-only searches, which have efficiencies of 15-25\% \citep[e.g.][]{ckr+15,cck+16}. 

This approach is meant to be complementary to standard search methods. Current radio pulsation searches preselect \fermi LAT unassociated sources based on a posteriori knowledge of pulsar properties \citep{kcj+12,ckr+15,hph+15}.} Criteria require that the gamma-ray sources are non-variable and that they have pulsar-like spectra (i.e. power laws with high-energy exponential cutoffs). Practical considerations also require that the position uncertainty of the \fermi source is small enough to fit within a single beam of radio dishes. Typically this implies semi-major axes \adeg{0.1} or less, depending on the search frequency \citep{cck+16,drc+16}. A final selection criterion, common in the most recent radio pulsation searches, is to select only high latitude \fermi unassociated sources ($\vert{b}\vert>$\adeg{5}).  As the high yield of discoveries demonstrates, this is an optimal strategy for finding MSPs but it largely misses young pulsars with their smaller scale height confining them to the Galactic plane. While \fermi has detected many young gamma-ray pulsars in the plane, the majority have not been strong radio emitters \citep{aaa+13,cwp+16}. It has been argued by \citet{ckr+12} and others that the lack of radio pulsations among these unassociated sources may be due to more narrowly beamed radio emission pointed away from our line of sight \citep[see][]{rwjk17}. They further argue that the majority of young pulsars have already been found in past large area radio surveys.

The pulsar candidates here have been selected without regard to their high-energy properties, the sizes of the error ellipses or the sky distribution of the unassociated sources. Ignoring the high-energy properties, at least initially, maximizes our opportunity for identifying outliers; e.g. the pulsar whose power-law spectrum breaks outside the \fermi energy band, the glitching or transitional pulsar with a time-variable light curve \citep[e.g.][]{sah+14}. This means that the candidate list can provide a useful check on the efficacy of existing selection processes for pulsation searches. 

This method makes no pre-selection based on Galactic latitude. Of the 16 candidates in Table \ref{tab_candi}, eight have $\vert{b}\vert<$\adeg{5}. None of these candidates would be considered radio-quiet \citep{ckr+12}.  We can extrapolate their spectra to 1.4 GHz and predict phase-averaged flux densities at least a factor of ten or more above the formal definition of a radio quiet pulsar S$_{1.4}<$30 $\mu$Jy. These may be MSPs that have been biased against in radio pulsation search samples. We note that that the two new MSPs discovered from Paper I and the discovery of an MSP toward the Galactic center are all at low latitudes \citep{c+18,bdf+17}. Some of the remaining low-latitude candidates may also be young pulsars that have been missed in past Galactic plane searches. In these instances since we know their positions and their phase-averaged flux densities, they may be targeted for more extensive searches at higher frequencies where the effects of absorption, dispersion and scattering are much reduced. 

As noted above, we are not restricted to small error ellipses. Half of the candidates in Table \ref{tab:gamma} have semi-major axes larger than \adeg{0.1}. They would be prohibitively time intensive to search for pulsations. Identifying a compact, steep-spectrum candidate with a localization of a few arcseconds within these large error ellipses increases the pulsation search efficiency and it enables deep searches for other multi-wavelength counterparts. 

The gamma-ray properties of this sample are also illustrative. In column four of Table \ref{tab_candi} we give a ranking that each \fermi source is likely to be a pulsar, based on its gamma-ray spectral properties \citep[see \S\ref{pulse} and][]{ckr+15}. These classifications are based on visual inspection of the \fermi spectra looking for pulsar-like characteristics that include little excess low-energy emission, peak significance in the 1 to 5 GeV range, and a sharp high energy cutoff with minimal emission above 10 GeV. Among those candidates selected using the radio image-based method, there is a preponderance of sources whose gamma-ray spectra are rated as ambiguous, poor, or unlikely pulsar candidates. There appear to be three main reasons for these low rankings; low signal-to-noise spectra (i.e. low TS), significant emission above 10 GeV, and spectra that deviate very little from a strict power-law. For the two 3FGL MSPs, machine-learning algorithms have independently confirmed that these would be low-ranked pulsar candidates based on their gamma-ray properties alone {sxy+16}. We have taken the list of 3FGL unassociated sources from \citet{sxy+16} for which two independent classifiers mark them as pulsar candidates, removed all recent newly discovered gamma-ray pulsars, and rank-ordered them by classifier scores.  3FGL\,J1827.6$-$0846 does slightly better, falling in the bottom half (random forest) or the bottom third logistic regression) of candidates, while 3FGL\,J1901.5$-$0126 lies in the last decile. In both cases the algorithms classify these MSPs as young pulsars.

The discovery of new pulsars in sources with non-pulsar-like gamma-ray properties implies that the pulsar discovery space using \fermi is larger than previously thought. It may also indicate that the gamma-ray only selection process has biased the known population of gamma-ray pulsars, and that expanded search methods may broaden the range of physical parameters that currently describe the gamma-ray pulsar population. Finally, these results make clear that, while gamma-ray spectral and variability properties are certainly instrumental in determining quality pulsar candidates, multi-wavelength techniques provide an alternative, and hitherto underutilized source selection method. 

Although encouraging, this image-based method is not without its shortcomings. A number of problems were discussed in \S\ref{method} with using radio surveys for identifying pulsar candidates. Additional data, such as follow-up high resolution imaging can be used but false positives have not been entirely eliminated from this sample. We have used the ATCA and VLA imaging to eliminate extended HzRGs but with a typical resolution of \asec{5}, we are only sensitive to structures at high redshift larger than 25 kpc. Approximately 30\% of well-studied HzRGs have more concentrated morphologies \citep{prc+00}, i.e. their total angular extent is less than \asec{5}. 

Fortunately, existing and future metre and centimetre wavelength radio interferometers can improve on the deficiencies of this sky-surveys approach. The wide instantaneous bandwidths of today's interferometers can measure a spectral index at many different frequencies simultaneously without variability issues or systematic calibration errors between surveys \citep{rbo16}. This capability is nicely illustrated by the GLEAM spectra for \textsc{Fermi}\,J1843.8$-$3834 (Fig.\,\ref{fig:spectra}). The long baselines of GMRT, LOFAR and VLA can make wide-field images at arcsecond resolution, measuring compactness directly. Polarization can serve as an additional discriminant, and it may be possible to identify candidates based on their diffractive scintillations as measured in variance images \citep{djb+16}. Even with these improvements, image-plane searches will never supplant direct pulsation searches nor are they meant to. This complementary approach should be used in special directions where enhanced scattering may be expected, such as the GeV excess or the Galactic plane in general.  Alternatively, it may be worthwhile to image those \fermi sources ranked as highly significant pulsar candidates \citep{sxy+16,mcf+16} but with previous (unsuccessful) pulsation searches, in case a tight binary or similarly exotic pulsar is hiding among the sample.

\section*{Acknowledgments}

We thank the staff at ATCA and the VLA for their generous allocation of Director's Discretionary Time. The National Radio Astronomy Observatory is a facility of the National Science Foundation operated under cooperative agreement by Associated Universities, Inc. The Australia Telescope Compact Array is part of the Australia Telescope National Facility which is funded by the Australian Government for operation as a National Facility managed by CSIRO. \textit{Fermi} work at NRL is supported by NASA. The \textit{Fermi} LAT Collaboration acknowledges generous ongoing support from a number of agencies and institutes that have supported both the development and the operation of the LAT as well as scientific data analysis. This work was performed in part under DOE Contract DE-AC02-76SF00515. Additionally, the LAT includes support from the National Aeronautics and Space Administration and the Department of Energy in the United States, the Commissariat \`a l'Energie Atomique and the Centre National de la Recherche Scientifique / Institut National de Physique Nucl\'eaire et de Physique des Particules in France, the Agenzia Spaziale Italiana and the Istituto Nazionale di Fisica Nucleare in Italy, the Ministry of Education, Culture, Sports, Science and Technology (MEXT), High Energy Accelerator Research Organization (KEK) and Japan Aerospace Exploration Agency (JAXA) in Japan, and the K.~A.~Wallenberg Foundation, the Swedish Research Council and the Swedish National Space Board in Sweden. Additional support for science analysis during the operations phase is gratefully acknowledged from the Istituto Nazionale di Astrofisica in Italy and the Centre National d'\'Etudes Spatiales in France. This research has made use of NASA's Astrophysics Data System Bibliographic Services. This work has also made extensive use of the SIMBAD and Vizier databases maintained by the Centre de Donn\'ees astronomiques de Strasbourg. 

{\it Facilities:} ATCA, \textit{Fermi} (LAT), VLA.

% make sure the AGN figure is put in the right place in the final submitted manuscript
%\clearpage
%\begin{figure}
%\includegraphics[width=7.5in]{image_for_paper.pdf}
%\caption{Contour plots for steep-spectrum radio candidates (left to right: 3FGL J1925.4+1727, 3FGL J1949.3+2433, 3FGL J2028.5+4040c, \textsc{Fermi}\,J2259.1$+$6233) in Fermi error ellipses that are resolved by the VLA. These are no longer considered pulsar candidates. The synthesized beam for each cutout is shown as a grey ellipse. The contour levels are at 3, 5, 7, 10, 12, 15, 20, 50 and 100 times the rms noise for the 3FGL sources, and 15, 20, 25, 40, 50, 100 times the rms noise for \textsc{Fermi}\,J2259.1$+$6233. The RMS noise in the respective images (left to right) is 48, 42, 44, and \mujybeam{23}. The center frequencies for the images lie between 2.8 GHz and 3 GHz. See \S\ref{obs:vla} and \S\ref{obs:atca} for details.}
%\label{fig:agn}
%\end{figure}

\end{document}